\documentclass[10pt,journal,twocolumn]{IEEEtran}

\usepackage{textcomp}
\usepackage{cite}
\usepackage{stfloats}
\usepackage{csquotes}
\usepackage{setspace}
\usepackage{graphicx,amsmath,amsfonts,amsthm,color,comment}
\usepackage{enumitem}
\usepackage{enumerate}
\usepackage{algorithmic}
\usepackage{pgfplots}
\usepackage{pgf}
\usepackage{tikz}
\usetikzlibrary{positioning,shapes}
\usepackage{amssymb}
\usepackage{csquotes}
\usepackage{bbm}
\usepackage{multirow}
\usepackage{bm}
\usepackage[nolist]{acronym}
\usepackage{amsthm}
\usepackage{arydshln}
\usepackage{subcaption}
\theoremstyle{definition}

\DeclareMathOperator*{\argmax}{arg\,max}

\usepackage[ruled,linesnumbered,vlined]{algorithm2e}
\SetNlSty{bfseries}{\color{black}}{}
\SetArgSty{textnormal}
\pgfplotsset{compat=1.15}
\SetKw{Continue}{continue}
\SetKw{Break}{break}

\begin{document}
\title{Soft-Output Successive Cancellation List Decoding}

\author{Peihong~Yuan,~\IEEEmembership{Member,~IEEE,}
        Ken~R.~Duffy,~\IEEEmembership{Senior Member,~IEEE,}
        Muriel M\'edard,~\IEEEmembership{Fellow,~IEEE,}

\thanks{This paper was presented in part at 2024 IEEE Int. Symp. Inf. Theory (ISIT)~\cite{yuan2024near}.}
\thanks{P. Yuan and M. M{\'e}dard are with the Massachusetts Institute of Technology Network Coding \& Reliable Communications Group (e-mails: \{phyuan, medard\}@mit.edu).}
\thanks{K. R. Duffy is with the Northeastern University Engineering Probability Information \& Communications Laboratory (e-mail: k.duffy@northeastern.edu).}
\thanks{This work was supported by the Defense Advanced Research Projects Agency (DARPA) under Grant HR00112120008. \emph{(Corresponding author: Muriel M{\'e}dard)}}
}
\begin{acronym}
    \acro{BCH}{Bose-Chaudhuri-Hocquenghem}
    \acro{GRAND}{guessing random additive noise decoding}
    \acro{SGRAND}{soft GRAND}
    \acro{DSGRAND}{discretized soft GRAND}
    \acro{SRGRAND}{symbol reliability GRAND}
    \acro{ORBGRAND}{ordered reliability bits GRAND}
    \acro{5G}{the $5$-th generation wireless system}
    \acro{APP}{a-posteriori probability}
    \acro{ARQ}{automated repeat request}
    \acro{ASK}{amplitude-shift keying}
    \acro{AWGN}{additive white Gaussian noise}
    \acro{B-DMC}{binary-input discrete memoryless channel}
    \acro{BEC}{binary erasure channel}
    \acro{BER}{bit error rate}
    \acro{biAWGN}{binary-input additive white Gaussian noise}
    \acro{BLER}{block error rate}
    \acro{UER}{undetected error rate}
    \acro{LER}{list error rate}
    \acro{bpcu}{bits per channel use}
    \acro{BPSK}{binary phase-shift keying}
    \acro{BSC}{binary symmetric channel}
    \acro{BSS}{binary symmetric source}
    \acro{CDF}{cumulative distribution function}
    \acro{CSI}{channel state information}
    \acro{CRC}{cyclic redundancy check}
    \acro{DE}{density evolution}
    \acro{DMC}{discrete memoryless channel}
    \acro{DMS}{discrete memoryless source}
    \acro{BMS}{binary input memoryless symmetric}
	\acro{eMBB}{enhanced mobile broadband}
	\acro{FER}{frame error rate}
	\acro{uFER}{undetected frame error rate}
	\acro{FHT}{fast Hadamard transform}
	\acro{GF}{Galois field}
	\acro{HARQ}{hybrid automated repeat request}
	\acro{i.i.d.}{independent and identically distributed}
	\acro{LDPC}{low-density parity-check}
    \acro{GLDPC}{generalized low-density parity-check}
	\acro{LHS}{left hand side}
	\acro{LLR}{log-likelihood ratio}
	\acro{MAP}{maximum-a-posteriori}
	\acro{MC}{Monte Carlo}
	\acro{ML}{maximum-likelihood}
	\acro{PDF}{probability density function}
	\acro{PMF}{probability mass function}
	\acro{QAM}{quadrature amplitude modulation}
	\acro{QPSK}{quadrature phase-shift keying}
	\acro{RCU}{random-coding union}
	\acro{RHS}{right hand side}
	\acro{RM}{Reed-Muller}
	\acro{RV}{random variable}
	\acro{RS}{Reed–Solomon}
	\acro{SCL}{successive cancellation list}
	\acro{SE}{spectral efficiency}
	\acro{SNR}{signal-to-noise ratio}
	\acro{UB}{union bound}
	\acro{BP}{belief propagation}
	\acro{NR}{new radio}
	\acro{CA-SCL}{CRC-assisted successive cancellation list}
	\acro{DP}{dynamic programming}
	\acro{URLLC}{ultra-reliable low-latency communication}
    \acro{MAC}{multiple access channel}
    \acro{SIC}{successive interference cancellation}
    \acro{RLNC}{random linear network coding}
    \acro{SINR}{signal-to-interference-plus-noise ratio}
    \acro{MDR}{misdetection rate}
    \acro{SC}{successive cancellation}
    \acro{SCL}{successive cancellation list}
    \acro{SO-SCL}{soft-output SCL}
    \acro{PM}{path metric}
    \acro{SISO}{soft-input soft-output}
    \acro{SO}{soft-output}
    \acro{SOGRAND}{soft-output GRAND}
    \acro{RM}{Reed-Muller}
    \acro{PM}{path metric}
    \acro{BCJR}{Bahl, Cocke, Jelinek and Raviv}
    \acro{MIMO}{multi-input multi-output}
    \acro{BICM}{bit-interleaved coded modulation}
    \acro{BICM-ID}{bit-interleaved coded modulation with iterative decoding}
    \acro{TUB}{truncated union bound}
    \acro{MCDE}{Monte Carlo density evolution}
    \acro{GMI}{generalized mutual information}
    \acro{SCAN}{soft cancellation}
    \acro{CN}{check node}
    \acro{VN}{variable node}
    \acro{MP}{message passing}
    \acro{SCS}{SC stack}
    \acro{SCOS}{SC ordered search}
    \acro{PAC}{polarization-adjusted convolutional}
\end{acronym}
\maketitle
\begin{abstract}
We introduce an algorithm for approximating the codebook probability that is compatible with all successive cancellation (SC)-based decoding algorithms, including SC list (SCL) decoding. This approximation is based on an auxiliary distribution that mimics the dynamics of decoding algorithms with an SC decoding schedule. Based on this codebook probability and SCL decoding, we introduce soft-output SCL (SO-SCL) to generate both blockwise and bitwise soft-output (SO).

Using that blockwise SO, we first establish that, in terms of both block error rate (BLER) and undetected error rate (UER), SO-SCL decoding of dynamic Reed-Muller (RM) codes significantly outperforms the CRC-concatenated polar codes from 5G New Radio under SCL decoding. Moreover, using SO-SCL, the decoding misdetection rate (MDR) can be constrained to not exceed any predefined value, making it suitable for practical systems.

Proposed bitwise SO can be readily generated from blockwise SO via a weighted sum of beliefs that includes a term where SO is weighted by the codebook probability, resulting in a soft-input soft-output (SISO) decoder. Simulation results for SO-SCL iterative decoding of product codes and generalized LDPC (GLDPC) codes, along with information-theoretical analysis, demonstrate significant superiority over existing list-max and list-sum approximations.
\end{abstract}

\begin{IEEEkeywords}
Polar coding, codebook probability, generalized decoding, joint error correction and detection, soft-input soft-output (SISO) decoding.
\end{IEEEkeywords}
\markboth
	{}
	{}

\section{Introduction}
Reliability in physical layer communication hinges on the frequency of forward error correction decoding errors. Undetected errors occur when the decoder provides a codeword that is distinct from the transmitted one and the system remains unaware of this erroneous decision. Undetected errors can be more harmful than detected errors, which are usually labeled as erasures. Consequently, code and decoder design objectives encompass not only reducing the \ac{BLER} but also maintaining a low \ac{UER}.

Decoding algorithms can be divided into two classes: complete and incomplete. Complete decoders always return a valid codeword and any \ac{ML} decoding algorithm essentially belongs to this group. In contrast, an incomplete decoder may provide estimates not fulfilling the conditions of being a member of the underlying code~\cite[Ch.1]{blahut2003algebraic}. If that occurs, the receiver is able to detect the error and, for  instance, request a retransmission. Both the \ac{BCJR}~\cite{bahl1974optimal} and \ac{BP} decoding algorithms are well-established incomplete decoders as a result of their focus on making bitwise decisions for the coded bits.

The standard method to convert a complete decoder into an incomplete one is to employ a \ac{CRC} outer code, which provides error detection capability after using a complete decoder for the inner code. The addition of the \ac{CRC} results in a reduced code-rate, and so the \ac{CRC} should be carefully designed to optimize the trade-off between the \ac{BLER} and \ac{UER}. The notion of an optimal incomplete decoding algorithm, introduced in~\cite{forney1968exponential}, can be viewed as implementing \ac{ML} decoding followed by a post-decoding threshold test that determines whether to accept or reject the \ac{ML} decision. This approach is optimal in the sense that there is no other decoding rule that simultaneously gives a lower \ac{BLER} and a lower \ac{UER}. The metric for evaluating this test can be efficiently carried out for terminated convolutional codes~\cite{raghavan1998reliability,hof2009optimal} and well approximated for tail-biting convolutional codes~\cite{hof2010performance} via a modification to the \ac{BCJR} algorithm. 

CRC-concatenated polar codes, as described in~\cite{tal2015list,niu2012crc}, result from the serial concatenation of polar codes with outer CRC codes. \Ac{SCL} decoding~\cite{tal2015list} is typically used to decode CRC-concatenated polar codes. First, a \ac{SCL} decoder creates a list of candidate decodings based on the inner polar code. If none of the candidates in the list pass the CRC test, a decoding failure is declared, i.e, an error is detected. Otherwise, the most likely candidate in the list is selected as the final decision, leading to an undetected error if it is not the same as the transmitted message. 

Optimal \ac{SISO} \ac{BCJR} decoding of general linear block codes requires complexity that is exponential in the number of redundancy bits. Pyndiah proposed a low complexity algorithm~\cite{pyndiah_1998} to extract the approximated bitwise \ac{SO} from a candidate list through list decoding. This approximation is used for parallel concatenated polar codes~\cite{liu2017parallel} and product polar codes~\cite{bioglio2019construction,condo2020practical,cocskun2024precoded} based on \ac{SCL} decoding. \Ac{BP}~\cite{arikan2010polar} and \ac{SCAN}~\cite{fayyaz2014low} decoding of polar codes provide bitwise \ac{SO} by applying the \ac{MP} algorithm on the polar encoding graph, using flooding-like and \ac{SC}-like schedules, respectively. Soft list decoding~\cite{xiang2020soft} starts with an \ac{SCL} decoding, finds the most likely candidate in the list and performs backwards \ac{BP} to generate the bitwise \ac{SO}.

Recently, a blockwise \ac{SO} measure has been developed for all \ac{GRAND} algorithms, e.g. \cite{Duffy19,duffy22ORBGRAND}, which takes the form of an accurate estimate of the \ac{APP} that a single decoding is correct or, in the case of list decoding, the probability that each element of the list is the transmitted codeword or the codeword is not contained in the list~\cite{galligan2023upgrade}. Core to the accuracy of the measure is the approximation that all unidentified codewords are uniformly distributed amongst unexplored sequences. In~\cite{yuan2023soft}, the bitwise \ac{SO} is approximated by utilizing the blockwise \ac{SO} to dynamically adjust the weight between list observation and channel observation.

In this study, we focus on the blockwise and bitwise \ac{SO} for polar~\cite{Arikan09} and polar-like codes. The novel contributions of this paper are summarised as follows.
\begin{itemize}
    \item\emph{Codebook probability of polar-like codes}: By extending the main idea in~\cite{galligan2023upgrade} from a guessing-based search to \ac{SC}-based tree search, we introduce an approximation of the codebook probability, which is the sum of the probability of all valid codewords given the channel observations.
    \item\emph{\Ac{SO-SCL} decoding}: Based on the codebook probability and \ac{SCL} decoding, we introduced \ac{SO-SCL} to generate both blockwise \ac{SO} and bitwise \ac{SO}.
    \item\emph{Generalized decoding with \ac{SO-SCL}}: The blockwise \ac{SO} generated by \ac{SO-SCL} accurately matches the probability of the output decision being the transmitted codeword. By availing of that blockwise \ac{SO}, we demonstrate that dynamic \ac{RM} codes using generalized decoding significantly outperform CRC-concatenated polar codes using \ac{SCL} decoding in terms of both \ac{BLER} and \ac{UER}. Furthermore, the \ac{MDR} can be constrained to not exceed any predefined value.
    \item\emph{Iterative decoding with \ac{SO-SCL}}: Information theoretic and simulation results of iterative decoding for product and GLDPC codes demonstrate the superiority of bitwise \ac{SO} generated by \ac{SO-SCL}.
\end{itemize}

This paper is organized as follows. Section~\ref{sec:prelim} provides background on the problem. An approximation of the codebook probability of polar-like codes, along with blockwise and bitwise \ac{SO} derived from the codebook probability, is developed in Section~\ref{sec:codebook_prob}. Section~\ref{sec:numerical} presents numerical results demonstrating the accuracy of the blockwise \ac{SO}, the performance of generalized decoding based on the blockwise \ac{SO}, the iterative decoding performance using the bitwise \ac{SO}, and the information-theoretic consideration of the bitwise \ac{SO}. Section~\ref{sec:conclusions} concludes the paper.

\section{Preliminaries}\label{sec:prelim}
\subsection{Notations}
In this paper, length-$N$ vectors are denoted as $x^N = \left(x_1, x_2, \dots, x_N\right)$, where we write $x_i$ for its $i$-th entry. For completeness, note that $x^0$ is void. A \ac{RV} is denoted by an uppercase letter, such as $X$, and its counterpart, e.g., $x$, is used for a realization. Then, a random vector is expressed as $X^N = \left(X_1, X_2, \dots, X_N\right)$. The \ac{PDF} of a continuous \ac{RV} and the \ac{PMF} of a discrete \ac{RV} evaluated at $x$ are denoted as $p_X(x)$, where the extensions to the vectors is straightforward. A \ac{B-DMC} is characterized by conditional probabilities $p_{Y|C}$, where the input takes on values in binary alphabet $\{0,1\}$ and the output set $\mathcal{Y}$ is specified by the considered channel model. For natural numbers, we write $[a] = \{i : i \in \mathbb{N}, 1 \leq i \leq a\}$.

\subsection{Codeword Probability and Codebook Probability}
For a binary linear block code, the codebook $\mathcal{C}$ contains all valid codewords. The codeword $c^N\in\mathcal{C}$ is transmitted via $N$ independent uses of a \ac{B-DMC} $P_{Y|C}$. 

To avoid ambiguities, we define $Q_{C^N|Y^N}(b^N|y^N)$ as the auxiliary conditional probability of sequence $b^N$ conditioned on channel observation $y^N$, which is not aware of the codebook information, i.e.,
\begin{align*}
Q_{C^N|Y^N}\left(b^N|y^N\right) \triangleq \prod_{i=1}^N P_{C|Y}(b_i|y_i),
\end{align*}
where $P_{C|Y}(b_i|y_i)$ is given by
\begin{align*}
P_{C|Y}(b_i|y_i) = \frac{P_{Y|C}(y_i|b_i)\cdot P_{C}(b_i)}{\sum_{a\in\left\{0,1\right\}}P_{Y|C}(y_i|a)\cdot P_{C}(a)}.
\end{align*}
On the other hand, the standard codeword probability conditioned on the channel observation $P_{C^N|Y^N}\left(b^N|y^N\right)$ is aware of the codebook information, i.e.,
\begin{align*}
&P_{C^N|Y^N}\left(b^N|y^N\right)=0,~\forall b^N\notin\mathcal{C}\\
&\sum_{b^N\in\mathcal{C}}P_{C^N|Y^N}(b^N|y^N)=1.
\end{align*}
The \emph{codebook probability} is defined as the sum of the auxiliary conditional probabilities of all valid codewords,
\begin{align}\label{eq:codebook_probability_c}
Q_\mathcal{C}\left(y^N\right)\triangleq \sum_{b^N\in \mathcal{C}}Q_{C^N|Y^N}\left(b^N|y^N\right).
\end{align}
Note that we have the relationship
\begin{align*}
P_{C^N|Y^N}\left(b^N|y^N\right) =\frac{Q_{C^N|Y^N}\left(b^N|y^N\right)}{ Q_\mathcal{C}\left(y^N\right)},~\forall b^N\in\mathcal{C}.
\end{align*}

\subsection{Polar-like Codes and Their Decoding}\label{sec:polar}
A binary polar-like code of block length $N$ and dimension $K$ is defined by a set $\mathcal{A}\subseteq \left[N\right]$ of indices with $|\mathcal{A}|=K$ and a set of linear functions $f_i$, $i\in\mathcal{F}$, where $N$ is a positive-integer power of $2$ and $\mathcal{F}\triangleq\left[N\right]\setminus \mathcal{A}$. The $K$-bit message is mapped onto the subvector $u_\mathcal{A}$ of the input $u^N$ to the polar transform, where the frozen bits are evaluated as
\begin{align}
u_i=f_i\left(u^{i-1}\right), \forall i\in\mathcal{F}.\label{eq:dyn_frozen_def}
\end{align}
Observe that each frozen bit $u_i$ is either statically set to zero (since the $f_i$s are linear) or they change according to the input $u^{i-1}$, which are called dynamic frozen bits~\cite{trifonov2015polar}. This representation unifies various modifications of polar codes, e.g., \ac{CRC}-concatenated  polar codes~\cite{tal2015list}, polar subcodes~\cite{trifonov2015polar} \ac{PAC} codes~\cite{arikan2019sequential} and dynamic \ac{RM} codes~\cite{cocskun2020successive}. The codeword is then obtained by applying a polar transform as follows
\begin{align}
c^N=u^N\mathbb{F}^{\otimes \log_2 N},
\end{align}
where $\mathbb{F}$ denotes the binary Hadamard matrix~\cite{Arikan09}. We define a set $\mathcal{U}$ that contains all decoding paths $u^N$ that satisfy the frozen constraints Eq.~\eqref{eq:dyn_frozen_def},
\begin{align*}
\mathcal{U}\triangleq \left\{u^N\in\left\{0,1\right\}^N: u_i=f_i\left(u^{i-1}\right), \forall i\in\mathcal{F} \right\}.
\end{align*}
Obviously, the codebook of polar codes is given by
\begin{align*}
\mathcal{C}=\left\{c^N\in\left\{0,1\right\}^N: c^N=u^N\mathbb{F}^{\otimes \log_2 N}, \forall u^N\in\mathcal{U}\right\}.
\end{align*}

At the receiver side, 
\ac{SC} decoding observes the channel output $y^N$ and performs a sequential greedy search to obtain decisions as follows
\begin{align}
\hat{u}_i=\left\{\begin{aligned}
& f_i\left(\hat{u}^{i-1}\right),~&i\in\mathcal{F}\\
& \argmax_{u\in\left\{0,1\right\}}Q_{U_i|Y^NU^{i-1}}\left(u|y^N\hat{u}^{i-1}\right),~&i\in\mathcal{A} \label{eq:SC_decoding}
\end{aligned}\right.
\end{align}
where $Q_{U^N|Y^N}$ denotes an auxiliary conditional \ac{PMF} induced by assuming that $U^N$ is uniformly distributed in $\{0,1\}^N$. This implies that $Q_{U^N|Y^N}$ assumes the frozen bits $U_\mathcal{F}$ to be also uniformly distributed and independent of the message bits $U_{\mathcal{A}}$. Observe that \ac{SC} decoding computes $Q_{U_i|Y^NU^{i-1}}\left(u_i|y^N\hat{u}^{i-1}\right)$ by treating $U_i$ and all upcoming frozen bits, namely $U_{i+1},\dots,U_N$, as uniformly distributed given the channel observation $y^N$ and previous decisions $\hat{u}^{i-1}$. In other words, the frozen constraints are used to determine which decision to make but don't impact the reliability of the decision. Then, a block error is declared only if $u_{\mathcal{A}}\neq \hat{u}_{\mathcal{A}}$ since Eq.~\eqref{eq:dyn_frozen_def} is already used in decoding via Eq.~\eqref{eq:SC_decoding}.

\ac{SCL} decoding~\cite{tal2015list,balatsoukas2015llr} tracks several \ac{SC} decoding paths in parallel. At each decoding phase $i\in\mathcal{A}$, instead of making a hard decision on $u_i$, two possible decoding paths are continued in parallel threads. The maximum number $2^{K}$ of paths implements \ac{ML} decoding but with exponential complexity in $K$. To limit complexity, one may keep up to $L$ paths at each phase. The reliability of partial decoding path $v^{i}$ is given by 
\begin{align}\label{eq:PM}
&Q_{U^i|Y^N}\left(v^i\left|y^N\right.\right)\nonumber\\
&=Q_{U^{i-1}|Y^N}\left(v^{i-1}\left|y^N\right.\right)Q_{U_i|Y^NU^{i-1}}\left(v_i\left|y^Nv^{i-1}\right.\right)
\end{align}
where the right-most term $Q_{U_i|Y^NU^{i-1}}\left(\hat{u}_i\left|y^N\hat{u}^{i-1}\right.\right)$ can be efficiently computed by the standard \ac{SC} decoding and $Q_{U^0|Y^N}\left(\varnothing|y^N\right)\triangleq 1$ by definition.\footnote{The term $-\log Q_{U^i|Y^N}\left(\hat{u}^i\left|y^N\right.\right)$ is called \ac{PM} for SC-based decoding in \ac{LLR} domain~\cite{balatsoukas2015llr}.} Note that the frozen constraints are used to determine which decision to make at frozen positions behaving as anchors and are irrelevant to the reliability of the decoding path. At the end of the $N$-th decoding phase, a list $\mathcal{L}_U$ of paths is collected.\footnote{In this work,  $\mathcal{L}_U$ is associated with the list of candidate decisions for $u^N$, while $\mathcal{L}_C$ is associated with the list of candidate decisions for codeword $c^N$, i.e., $\mathcal{L}_C\triangleq \left\{c^N\in\left\{0,1\right\}^N: c^N=u^N\mathbb{F}^{\otimes \log_2 N}, \forall u^N\in\mathcal{L}_U \right\}$.}  Finally, the output is the decoding path maximizing the path reliability:
\begin{equation*}
    \hat{u}^N = \argmax_{v^{N}\in\mathcal{L}_U} Q_{U^N|Y^N}\left(v^N|y^N\right).
\end{equation*}
In addition to SCL decoders, there are other improved SC-based decoders for polar-like codes such as \ac{SCS}~\cite{niu2012stack,miloslavskaya2014sequential}, SC-Fano~\cite{jeong2019sc,arikan2019sequential} and \ac{SCOS}~\cite{yuan2021complexity,yuan2024successive}, which employ different search strategies on the SC-decoding tree, utilizing the reliability measure in Eq.~\eqref{eq:PM}.

\subsection{Forney's Generalized Decoding}\label{sec:Gdecoding}
Forney introduced a generalized decoding rule~\cite{forney1968exponential}, which relies on a threshold test. The decoder output $\hat{c}^N$ is accepted if 
\begin{align}\label{eq:forney}
\frac{p_{Y^N|C^N}\left(y^N\left|\hat{c}^N\right.\right)}{\sum_{c^N\in \mathcal{C}} p_{Y^N|C^N}\left(y^N\left|c^N\right.\right) } \geq \frac{2^{NT}}{1+2^{NT}},
\end{align}
where the threshold parameter $T\geq 0$ controls the tradeoff between \ac{BLER} and \ac{UER}. Otherwise, the decision is rejected and decoder outputs an erasure flag, resulting in a detected error. Forney's generalized decoding rule is optimal in the sense of minimizing the \ac{UER} for a given \ac{BLER}
(and vice versa). Since the denominator of Eq.~\eqref{eq:forney} is difficult to compute in general, we may use a suboptimal decoding rule~\cite{forney1968exponential,hof2010performance,sauter2023error} based on list decoding with 
\begin{align*}
\!\!\!\frac{p_{Y^N|C^N}\left(y^N\left|\hat{c}^N\right.\right)}{\sum_{c^N\in \mathcal{C}} p_{Y^N|C^N}\left(y^N\left|c^N\right.\right) }\approx \frac{p_{Y^N|C^N}\left(y^N\left|\hat{c}^N\right.\right)}{\sum_{c^N\in \mathcal{L}_C} p_{Y^N|C^N}\left(y^N\left|c^N\right.\right) }.
\end{align*}
where $\mathcal{L}_C$ contains the candidate decisions of codeword $c^N$ obtained from the list decoding. Clearly, the approximation is precise when the list decoder exhaustively enumerates the entire codebook.

\subsection{SISO Decoding}\label{sec:SISO}
In various applications, the system requires post-decoding bitwise \ac{SO}, e.g., iterative detection and decoding of \ac{MIMO} system, \ac{BICM-ID}, product codes~\cite{elias_error-free_1954} and \ac{GLDPC} codes~\cite{liva2008quasi,Lentmaier10}. A SISO decoder takes the sum of channel \acp{LLR} $\ell_{\text{ch},i}$, and a-priori \acp{LLR}, denoted as $\ell_{\text{A},i}$, as input,
\begin{align*}
\ell_{\text{ch},i} \triangleq\log\frac{p_{Y|C}(y_i|0)}{p_{Y|C}(y_i|1)},~
\ell_{\text{A},i} \triangleq \log\frac{P_{C_i}(0)}{P_{C_i}(1)},~i\in\left[N\right].
\end{align*}
An optimal \ac{SISO} decoder outputs \ac{APP} \acp{LLR}, represented as $\ell_{\text{APP},i}$, and extrinsic \acp{LLR}, represented as $\ell_{\text{E},i}$.
\begin{align*}
\ell_{\text{APP},i} &\triangleq \log\frac{P_{C_i|Y^N}\left(0\left|y^N\right.\right) }{P_{C_i|Y^N}\left(1\left|y^N\right.\right) }\\
&= \log \frac{\sum_{c_i=0,c^N\in\mathcal{C}} Q_{C^N|Y^N}\left(c^N\left|y^N\right.\right)}{\sum_{c_i=1,c^N\in\mathcal{C}} Q_{C^N|Y^N}\left(c^N\left|y^N\right.\right)}\\
\ell_{\text{E},i} &\triangleq \ell_{\text{APP},i} - \ell_{\text{A},i} - \ell_{\text{ch},i},~i\in\left[N\right].
\end{align*}
For an $(N, K)$ block code, the \ac{APP} \acp{LLR} can be determined using the \ac{BCJR} algorithm~\cite{bahl1974optimal} with $2^{N-K}$ states. 

Clearly, \ac{APP} \acp{LLR} can be approximated from a candidate list through list decoding by considering only the list $\mathcal{L}_C$ instead of the entire codebook $\mathcal{C}$ (this is called \emph{list-sum} approximation),
\begin{align}
\ell_{\text{APP},i}^\text{list-sum}=
      \log \frac{\sum_{c_i=0,c^N\in\mathcal{L}_C} Q_{C^N|Y^N}\left(c^N\left|y^N\right.\right)}{\sum_{c_i=1,c^N\in\mathcal{L}_C} Q_{C^N|Y^N}\left(c^N\left|y^N\right.\right)}.\label{eq:list-sum}
\end{align}
Pyndiah proposed a low complexity algorithm~\cite{pyndiah_1998} to approximate \ac{APP} \acp{LLR} from a list, which is generally deemed suitable for log-domain implementations (this is called \emph{list-max} approximation).
\begin{align}
\ell_{\text{APP},i}^\text{list-max}=
      \log \frac{\max_{c_i=0,c^N\in\mathcal{L}_C} Q_{C^N|Y^N}\left(c^N\left|y^N\right.\right)}{\max_{c_i=1,c^N\in\mathcal{L}_C} Q_{C^N|Y^N}\left(c^N\left|y^N\right.\right)}.\label{eq:list-max}
\end{align}
If there is no competing codeword in $\mathcal{L}_C$ for the $i$-th bit, a predefined saturation value $\beta$ is required for both list-sum and list-max approximation~\cite{pyndiah_1998}, i.e., 
\begin{align}
\ell_{\text{APP},i}^\text{list-max/sum}=\left\{
    \begin{aligned}
      &\ell_{\text{A},i} + \ell_{\text{ch},i}+\beta, \text{ if }\forall c^N\in\mathcal{L}_C, c_i=0\\
      &\ell_{\text{A},i} + \ell_{\text{ch},i}-\beta,\text{ if }\forall c^N\in\mathcal{L}_C, c_i=1.\\
    \end{aligned}\right.\label{eq:pyndiah_e}
\end{align}
 
\section{Codebook Probability of Polar-like Codes}\label{sec:codebook_prob}
Since the polar transform $c^N=u^N\mathbb{F}^{\otimes \log_2 N}$ is a one-to-one mapping, we have
\begin{align*}
Q_{C^N|Y^N}\left(c^N\left|y^N\right.\right)=Q_{U^N|Y^N}\left(u^N\left|y^N\right.\right).
\end{align*}
The codebook probability Eq.~\eqref{eq:codebook_probability_c} can be rewritten as the sum of probabilities for all valid decoding paths,
\begin{align}\label{eq:codebook_probability_u}
Q_\mathcal{U}\left(y^N\right)= \sum_{v^N\in \mathcal{U}}Q_{U^N|Y^N}\left(v^N|y^N\right).
\end{align}
To compute the exact value of $Q_\mathcal{U}\left(y^N\right)$, a full traversal of the SC decoding tree is required. 

Here we introduce a method to approximate the codebook probability of polar and polar-like codes by using SC-based decoding.
\begin{figure}[t]
	\centering
	\begin{tikzpicture}[scale=1.2]
    \footnotesize
\draw [dashed] (0.5,4)--(0.1,3); \draw [dashed]  (0.5,4)--(0.9,3);
\draw [thick] (2,4)--(1.6,3); \draw [dashed] (2,4)--(2.4,3);
\draw [dashed] (3.5,4)--(3.1,3); \draw [dashed] (3.5,4)--(3.9,3);
\draw [dashed] (5,4)--(4.6,3); \draw [dashed] (5,4)--(5.4,3);

\draw [dashed] (1.25,5)--(0.5,4); \draw [thick]  (1.25,5)--(2,4);
\draw [dashed] (4.25,5)--(3.5,4); \draw [dashed] (4.25,5)--(5,4); 

\draw [thick] (2.75,6)--(1.25,5); \draw[dashed] (2.75,6)--(4.25,5);

\draw[dotted] (-0.5,2) to (6,2);
\node at (-0.75,2.5) {message}; 
\draw[dotted] (-0.5,3) to (6,3);
\node at (-0.75,3.5) {frozen}; 
\draw[dotted] (-0.5,4) to (6,4);
\node at (-0.75,4.5) {message}; 
\draw[dotted] (-0.5,5) to (6,5);
\node at (-0.75,5.5) {frozen}; 
\draw[dotted] (-0.5,6) to (6,6);

\draw [dashed] (0.1,3)--(-0.1,2); \draw [dashed]  (0.1,3)--(0.3,2);
\draw [dashed] (0.9,3)--(0.7,2); \draw [dashed]  (0.9,3)--(1.1,2);
\draw [thick] (1.6,3)--(1.4,2); \draw [dashed]  (1.6,3)--(1.8,2);
\draw [dashed] (2.4,3)--(2.2,2); \draw [dashed]  (2.4,3)--(2.6,2);
\draw [dashed] (3.1,3)--(2.9,2); \draw [dashed]  (3.1,3)--(3.3,2);
\draw [dashed] (3.9,3)--(3.7,2); \draw [dashed]  (3.9,3)--(4.1,2);
\draw [dashed] (4.6,3)--(4.4,2); \draw [dashed]  (4.6,3)--(4.8,2);
\draw [dashed] (5.4,3)--(5.2,2); \draw [dashed]  (5.4,3)--(5.6,2);
\node at (2.75,6.3) {$Q_{U^0|Y^N}(\varnothing|y^N)=1$};
\draw[fill] (2.75,6) circle (1.5pt);
\node at (0.3,1) {visited leaf: $0100$};
\draw[fill] (1.4,2) circle (1.5pt);
\draw [shorten >=2pt,shorten <=6pt,->] ((0.3,1)--(1.4,1.9);

\draw[blue, dashed, rounded corners] (2.8,1.9) rectangle (5.7,5.5);
\draw[blue, fill] (4.25,5) circle (1.5pt);
\draw[blue, dashed, rounded corners] (2.1,1.9) rectangle (2.7,3.5);
\draw[blue, fill] (2.4,3) circle (1.5pt);

\draw[red, dashed, rounded corners] (1.6,1.9) rectangle (2.0,2.5);
\draw[red, fill] (1.8,2) circle (1.5pt);
\draw[red, dashed, rounded corners] (-0.2,1.9) rectangle (1.2,4.5);
\draw[red, fill] (0.5,4) circle (1.5pt);

\node at (5,1) {\textcolor{blue}{invalid subtrees}};
\draw [blue,shorten >=5pt,shorten <=6pt,->] ((5,1)--(4.25,1.9);
\draw [blue,shorten >=5pt,shorten <=10pt,->] ((4.8,1)--(2.4,1.9);

\node at (2.7,1) {\textcolor{red}{unvisited subtrees}};
\draw [red,shorten >=5pt,shorten <=6pt,->] ((2.8,1)--(1.8,1.9);
\draw [red,shorten >=5pt,shorten <=10pt,->] ((2.6,1)--(0.5,1.9);

\end{tikzpicture}
	\caption{Example of the SC decoding tree of a polar code with frozen bits $u_1=u_3=0$. The whole decoding tree consists of three parts: a) visited leaf: the SC output $\hat{u}^4=(0,1,0,0)$. b) invalid subtrees: the subtree rooted at $\hat{u}_1=1$ and the subtree rooted at $\hat{u}^3=(0,1,1)$. c) unvisited subtrees: the subtree rooted at $\hat{u}^2=(0,0)$ and the leaf $\hat{u}^4=(0,1,0,1)$.}
	\label{fig:sc_tree}
\end{figure}
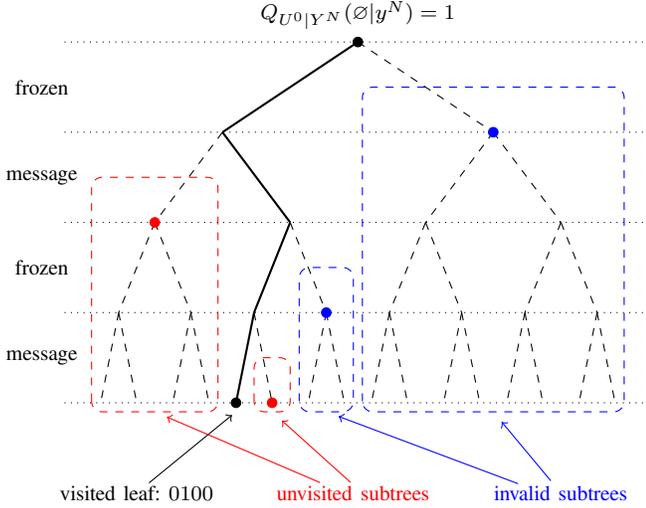
An SC-based decoding algorithm divides the decoding tree into three parts,
\begin{itemize}
    \item [a)] \emph{Visited leaves} denote the valid visited decoding paths of depth $N$. Note that we may have more than one visited leaf, e.g., a \ac{SCL} decoder returns $L$ visited leaves.
    \item [b)] \emph{Invalid subtrees} stands for the subtrees rooted at the nodes which are not visited during the decoding due to the conflict of frozen constraints.
    \item [c)] \emph{Unvisited subtrees} are the subtrees rooted at the nodes which satisfy the frozen constraints, but are not visited (usually due to the complexity issues).
\end{itemize}
Now we define sets $\mathcal{V}$, $\mathcal{W}$ and $\mathcal{I}$ containing the visited leaves, the roots of unvisited subtrees, and the roots of invalid subtrees, respectively. For the mini example with \ac{SC} decoding shown in Fig.~\ref{fig:sc_tree}, we have 
\begin{align*}
\mathcal{V} &= \left\{(0,1,0,0)\right\}\\
\mathcal{W} &= \left\{(0,0),(0,1,0,1)\right\}\\
\mathcal{I} &= \left\{(1),(0,1,1)\right\}.
\end{align*}
The codebook probability $Q_\mathcal{U}\left(y^N\right)$ is then written as 
\begin{align*}
\underbrace{\sum_{v^N\in \mathcal{V}}Q_{U^N|Y^N}\left(v^N|y^N\right)}_{\text{(a) all visited leaves}}+\overbrace{\sum_{a^i\in \mathcal{W}} \!\!\underbrace{\sum_{\substack{v^N\in \mathcal{U}\\v^i=a^i}}Q_{U^N|Y^N}\left(v^N|y^N\right)}_{\text{(c) all valid leaves underneath node $a^i$}}}^{\text{(b) all unvisited valid leaves}}.
\end{align*}
The term (c) describes the sum the of probabilities for all valid decoding paths (leaves) underneath node $a^i$.  By extending the approach~\cite[Cor.3]{galligan2023upgrade} from guess-based search to tree search, we assume that the leaves are uniformly distributed underneath the unvisited node $a^i$. We have the approximation 
\begin{align}\label{eq:core_approx}
\text{term (c)}\approx 2^{-\left|\mathcal{F}^{(i:N)}\right| }Q_{U^i|Y^N}\left(a^i|y^N\right),
\end{align}
where $\mathcal{F}^{(i:N)}$ denotes the set of indices for the frozen bits in the future, i.e., 
\begin{align*}
\mathcal{F}^{(i:N)} = \left\{j: j\in\mathcal{F}, i\leq j\leq N\right\},
\end{align*}
and $\left|\mathcal{F}^{(i:N)}\right|$ is the number of frozen bits in the future. The codebook probability Eq.~\eqref{eq:codebook_probability_u} is then approximated by
\begin{align}\label{eq:APU}
Q_\mathcal{U}\left(y^N\right)\approx~& Q_\mathcal{U}^*\left(y^N\right) \triangleq \underbrace{\sum_{v^N\in \mathcal{V}} Q_{U^N|Y^N}\left(v^N|y^N\right)}_{\substack{\text{sum of prob. for all visited leaves}}}\nonumber\\
&+\underbrace{\sum_{a^i\in \mathcal{W}} \!\! 2^{-\left|\mathcal{F}^{(i:N)}\right| }Q_{U^i|Y^N}\left(a^i|y^N\right)}_{\substack{\text{approx. sum of prob. for all unvisited valid leaves}}}.
\end{align}

The algorithm to compute $Q_\mathcal{U}^*\left(y^N\right)$ is compatible with all SC-based decoders. During SC-based decoding, when the decoder decides \emph{not} to visit a subtree rooted at the node $a^i$, $i\in\mathcal{A}$, we accumulate the probability $2^{-\left|\mathcal{F}^{(i:N)}\right|}Q_{U^i|Y^N}\left(a^i|y^N\right)$ as the approximated sum of probabilities for all valid leaves underneath node $a^i$. For the example in Fig.~\ref{fig:sc_tree}, we have 
\begin{align*}
Q_\mathcal{U}^*\left(y^4\right) &=Q_{U^4|Y^4}\left(0100|y^4\right) \\
&+ 2^{-1}\cdot Q_{U^2|Y^4}\left(00|y^4\right) + 2^{0}\cdot Q_{U^4|Y^4}\left(0101|y^4\right).
\end{align*}
Note that the terms $Q_{U^2|Y^4}\left(00|y^4\right)$ and $Q_{U^4|Y^4}\left(0101|y^4\right)$ are computed by the \ac{SC} decoder and do not lead to any additional computational complexity in the evaluation of the codebook probability.

\subsection{Blockwise Soft-Output}\label{sec:block_soft}
As in~\cite{galligan2023upgrade}, we define the probability of \emph{the output decision $\hat{u}^N$ being the transmitted codeword} as the (exact) blockwise \ac{SO} of the decision $\hat{u}^N$, i.e.,
\begin{align}\label{eq:Gamma}
\Gamma\left(y^N, \hat{u}^N\right)
\triangleq\frac{Q_{U^N|Y^N}\left(\hat{u}^N\left|y^N\right.\right)}{Q_\mathcal{U}\left(y^N\right)}
\end{align}
which is equivalent to the left hand side of Eq.~\eqref{eq:forney}. The blockwise \ac{SO} can be extended from a single decision to a list of candidates.
\begin{align}\label{eq:Gamma_list}
\Gamma\left(y^N, \mathcal{L}_U\right)
&\triangleq \frac{\sum_{v^N\in\mathcal{L}_U}Q_{U^N|Y^N}\left(v^N|y^N\right)}{Q_\mathcal{U}\left(y^N\right)},
\end{align}
which describes the probability of \emph{the candidate list $\mathcal{L}_U$ contains the transmitted codeword}.

Based on the blockwise \ac{SO}, we apply a threshold test for generalized decoding, i.e., joint error correction and detection,
\begin{align}\label{eq:decision_rule_Gamma}
\Gamma\left(y^N, \hat{u}^N\right) > 1-\epsilon.
\end{align}
The final decision $\hat{u}^N$ is accepted when Eq.~\eqref{eq:decision_rule_Gamma} is satisfied; otherwise, the decoder returns an erasure flag. As $\Gamma\left(y^N, \hat{u}^N\right)$ evaluates the probability of the output decision being correct, the decision rule mentioned above imposes an upper limit of $\epsilon$ on the \ac{MDR}, where \ac{MDR} is defined as the probability of an accepted decision being erroneous, i.e., the ratio between \ac{UER} and \ac{BLER}.

\subsection{Bitwise Soft-Outputs}\label{sec:bit_soft}
By using the same approach as described in~\cite{yuan2023soft}, we approximate the bitwise \ac{SO} $\ell_{\text{APP},i}$ and $\ell_{\text{E},i}$ based on $Q_\mathcal{U}^*\left(y^N\right)$ via Eq.~\eqref{eq:approx_app}. The approximated \ac{APP} \ac{LLR} $\ell^*_{\text{APP},i}$ introduces an additional term $\phi$ to dynamically adjust the weight between list observation and channel observation, where $\phi$ is the approximated sum of the probabilities for all codewords not in the list, i.e.,
\begin{align*}
\phi\approx\sum\limits_{c^N\in\mathcal{C},c^N\notin\mathcal{L}_C}Q_{C^N|Y^N}\left(c^N|y^N\right).
\end{align*}
If the candidates in the list are reliable (small $\phi$), $\ell^*_{\text{APP},i}$ is close to list-max approximation; otherwise (large $\phi$), $\ell^*_{\text{APP},i}$ is close to the decoder input $\ell_{\text{ch},i}+\ell_{\text{A},i}$. Furthermore, Eq.~\eqref{eq:approx_app} eliminates the need for the saturation value $\beta$ present in list-sum and list-max approximations Eq.~\eqref{eq:pyndiah_e}. 


\begin{table*}[ht]
\begin{equation}\label{eq:approx_app}
\begin{aligned}
\ell_{\text{APP},i}\approx\ell^*_{\text{APP},i}&= \log \frac{\sum\limits_{c_i=0,c^N\in\mathcal{L}_C} Q_{C^N|Y^N}\left(c^N\left|y^N\right.\right) +\phi\cdot P_{C|Y}\left(0\left|y_i\right.\right)}{\sum\limits_{c_i=1,c^N\in\mathcal{L}_C} Q_{C^N|Y^N}\left(c^N\left|y^N\right.\right)+ \phi\cdot P_{C|Y}\left(1\left|y_i\right.\right)}\\
\text{where }\phi&=Q_\mathcal{U}^*\left(y^N\right) - \!\!\!\sum\limits_{c^N\in\mathcal{L}_C}Q_{C^N|Y^N}\left(c^N|y^N\right)
\end{aligned}
\end{equation}
\end{table*}

\section{Numerical Results}\label{sec:numerical}
In this section, we present numerical results of \ac{SO-SCL}, demonstrating:
\begin{itemize}
\item the accuracy of the approximated blockwise \ac{SO},
\item the performance of generalized decoding using Eq.~\eqref{eq:decision_rule_Gamma},
\item the iterative decoding performance using Eq.~\eqref{eq:approx_app},
\item the quality of the bitwise \ac{SO} Eq.~\eqref{eq:approx_app}.
\end{itemize}
A large collection of polar-like encodings exist~\cite{mori2009performance,trifonov2012efficient,tal2013construct,mondelli2014polar,li2014rm,trifonov2015polar,wang2016parity,trifonov2017randomized,he2017beta,elkelesh2019decoder,yuan2019polar,cocskun2020successive,miloslavskaya2020design,bioglio2020design,miloslavskaya2021recursive,cocskun2022information,5gpolar}. As the design of polar-like codes is beyond the scope of our work, here we consider polar-like codes with two basic, channel-independent types of information sets,
\begin{itemize}
    \item 5G polar codes~\cite{5gpolar}: the information set is selected according to a reliability sequence,
   \item \ac{RM} codes~\cite{muller1954application,reed1954class}: the information set is selected according to the row weight in $\mathbb{F}^{\otimes \log_2 N}$~\cite{li2014rm},
\end{itemize}
and two types of frozen constraints,
\begin{itemize}
   \item static frozen bits: $u_i=0,~i\in\mathcal{F}$
   \item (convolutional) dynamic frozen bits~\cite{arikan2019sequential} of constraint length $6$:
   \begin{align*}
   u_i=u_{i-2}\oplus u_{i-3}\oplus u_{i-5}\oplus u_{i-6},~i\in\mathcal{F},i>6.
   \end{align*}
\end{itemize}
In general, 5G polar codes and \ac{RM} codes represent two extreme cases: 5G polar codes perform well under \ac{SC} decoding, whereas \ac{RM} codes perform well under \ac{ML} decoding. Dynamic frozen bits~\cite{trifonov2015polar} enhance the \ac{ML} performance of \emph{some} polar-like codes without compromising their \ac{SC} performance. 

\subsection{Accuracy of the approximated blockwise SO}\label{sec:accuracy_block}
By using the approximated codebook probability $Q_\mathcal{U}^*\left(y^N\right)$, we denote the approximated blockwise \ac{SO} for single decision and list of candidates by $\Gamma^*\left(y^N, \hat{u}^N\right)$ and $\Gamma^*\left(y^N, \mathcal{L}_U\right)$, respectively.,
\begin{align}\label{eq:approx_Gamma}
\Gamma^*\left(y^N, \hat{u}^N\right)
&\triangleq \frac{Q_{U^N|Y^N}\left(\hat{u}^N|y^N\right)}{Q_\mathcal{U}^*\left(y^N\right)}\\
\Gamma^*\left(y^N, \mathcal{L}_U\right)
&\triangleq \frac{\sum_{v^N\in\mathcal{L}_U}Q_{U^N|Y^N}\left(v^N|y^N\right)}{Q_\mathcal{U}^*\left(y^N\right)}.
\end{align}

To evaluate whether the approximated blockwise \ac{SO} $\Gamma^*\left(y^N, \hat{u}^N\right)$ matches the probability of the output decision being the correct codeword, we design a \ac{MC} simulation as follows. In the simulation, the codewords are transmitted over \ac{biAWGN} channels. The \ac{SO-SCL} decoder outputs a decision $\hat{u}^N$ and blockwise \ac{SO} $\Gamma^*\left(y^N, \hat{u}^N\right)$. We gather blocks with $1-\Gamma^*\left(y^N, \hat{u}^N\right)$ within specific ranges,
\begin{align*}
\left[1,10^{-0.5}\right),\left[10^{-0.5},10^{-1}\right),\dots,\left[10^{-4.5},10^{-5}\right)
\end{align*}
and compare their \ac{BLER} to $\text{E}\left[1-\Gamma^*\left(y^N, \hat{u}^N\right)\right]$ (solid lines). For reference, we also show the Forney's approximation (dashed lines) with list size $\left|\mathcal{L}_U\right|=L^\prime$,
\begin{align*}
\Gamma^\prime\left(y^N, \hat{u}^N\right)=\frac{Q_{U^N|Y^N}\left(\hat{u}^N|y^N\right)}{\sum_{u^N\in \mathcal{L}_U} Q_{U^N|Y^N}\left(u^N|y^N\right) }
\end{align*}
which approximates the codebook probability $Q_\mathcal{U}\left(y^N\right)$ as the list probability $\sum_{u^N\in \mathcal{L}_U}Q_{U^N|Y^N}\left(u^N|y^N\right)$.

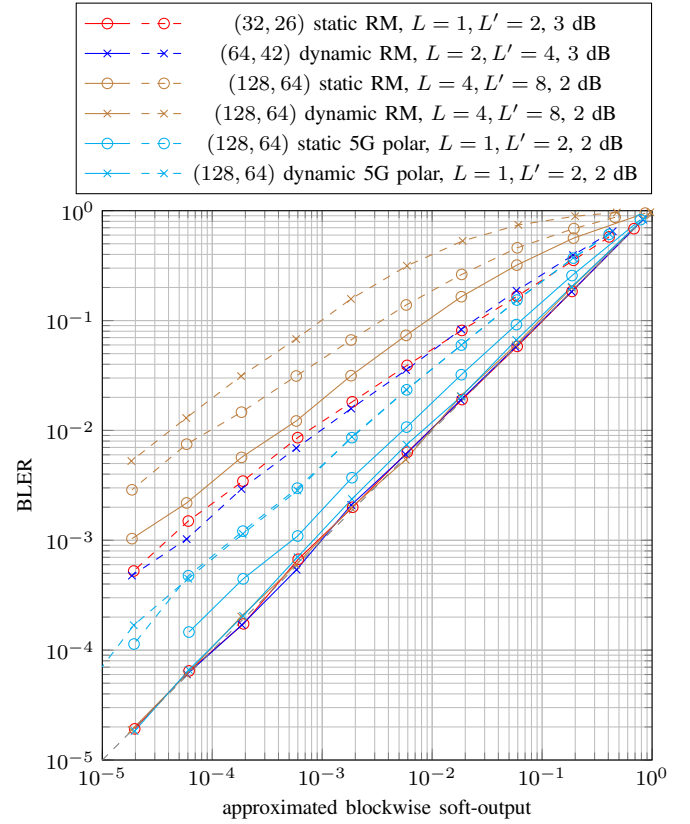
\begin{figure}[t]
	\centering
	\begin{tikzpicture}[scale=1]
\footnotesize
\begin{loglogaxis}[
legend style={at={(1,1.03)},anchor= south east},
legend columns=2,
ymin=1e-5,
ymax=1,
width=3.5in,
height=3.5in,
grid=both,
xmin = 1e-5,
xmax = 1,
xlabel = {approximated blockwise soft-output},
ylabel = {BLER},
]

\addplot[red,mark=o]
table[]{x y
0.68465 0.6862
0.18701 0.18418
0.059331 0.05821
0.01871 0.019093
0.0059424 0.0063431
0.0018932 0.002
0.00060471 0.00067245
0.00019266 0.00017354
6.1536e-05 6.4731e-05
1.9694e-05 1.9219e-05
};\addlegendentry{ }\label{line:ebch32_26}

\addplot[red,mark=o,dashed,mark options=solid]
table[]{x y
0.40717 0.57757
0.19278 0.35342
0.05926 0.16681
0.018616 0.081329
0.0059093 0.039094
0.0018762 0.018236
0.00059766 0.0085912
0.00019001 0.0034576
6.0495e-05 0.0014962
1.9302e-05 0.00052545
};\addlegendentry{$(32,26)$ static RM, $L=1,L^\prime=2$, $3$~dB}

\addplot[blue,mark=x]
table[]{x y
0.82841 0.83096
0.18517 0.18345
0.057298 0.057451
0.018139 0.018902
0.0057459 0.0060413
0.0018368 0.0020973
0.00058421 0.00053989
0.00018582 0.00016962
5.8782e-05 6.0295e-05
};\addlegendentry{ }\label{line:rm64_42}

\addplot[blue,mark=x,dashed,mark options=solid]
table[]{x y
0.43791 0.65001
0.19093 0.3937
0.058424 0.18695
0.018358 0.082943
0.0057986 0.035497
0.0018322 0.015809
0.00058147 0.0069023
0.00018406 0.0029298
5.8519e-05 0.001025
1.8584e-05 0.0004769
};\addlegendentry{$(64,42)$ dynamic RM, $L=2,L^\prime=4$, $3$~dB}

\addplot[brown,mark=o]
table[]{x y
0.86614 0.95106
0.1929 0.56292
0.058839 0.3197
0.018424 0.16486
0.0057879 0.073356
0.0018291 0.031521
0.00058014 0.012192
0.00018452 0.0056852
5.8469e-05 0.0021875
1.8601e-05 0.0010356
};\addlegendentry{ }\label{line:rm128_64}

\addplot[brown,mark=o,dashed,mark options=solid]
table[]{x y
0.46093 0.8709
0.19454 0.68739
0.059393 0.45953
0.018563 0.26302
0.0058086 0.13872
0.0018291 0.066569
0.00058016 0.03129
0.00018431 0.014665
5.8491e-05 0.0074959
1.8606e-05 0.0028797
};\addlegendentry{$(128,64)$ static RM, $L=4,L^\prime=8$, $2$~dB}

\addplot[brown,mark=x]
table[]{x y
0.96441 0.96751
0.18575 0.19933
0.057426 0.059787
0.018176 0.020496
0.0057436 0.005419
0.001827 0.0019762
0.00057762 0.0005991
0.00018355 0.0002025
5.8241e-05 6.0244e-05
1.8451e-05 1.8775e-05
};\addlegendentry{ }\label{line:drm128_64}

\addplot[brown,mark=x,dashed,mark options=solid]
table[]{x y
0.47733 0.9645
0.19976 0.89098
0.060849 0.74245
0.018929 0.53091
0.0058994 0.31508
0.0018271 0.15733
0.00057904 0.067932
0.00018272 0.031228
5.788e-05 0.012988
1.837e-05 0.0052696
};\addlegendentry{$(128,64)$ dynamic RM, $L=4,L^\prime=8$, $2$~dB}

\addplot[cyan,mark=o]
table[]{x y
0.78461 0.82842
0.18693 0.25671
0.058366 0.091953
0.018415 0.032156
0.005852 0.010721
0.0018652 0.003713
0.00059798 0.0010943
0.00019101 0.00044612
6.1345e-05 0.00014616
};\addlegendentry{ }\label{line:polar128_64}

\addplot[cyan,mark=o,dashed,mark options=solid]
table[]{x y
0.4073 0.60965
0.19224 0.37051
0.058939 0.15377
0.018446 0.05985
0.0058588 0.02343
0.0018616 0.0085916
0.00059404 0.0029832
0.00018988 0.0012113
6.0738e-05 0.00047601
1.9423e-05 0.00011367
};\addlegendentry{$(128,64)$ static 5G polar, $L=1,L^\prime=2$, $2$~dB}

\addplot[cyan,mark=x]
table[]{x y
0.82749 0.83664
0.18583 0.20332
0.058206 0.066571
0.018397 0.020398
0.0058511 0.0074201
0.0018632 0.0023738
0.00059616 0.00070268
0.00019062 0.00020486
6.1184e-05 6.6146e-05
1.9661e-05 1.8353e-05
};\addlegendentry{ }\label{line:dpolar128_64}

\addplot[cyan,mark=x,dashed,mark options=solid]
table[]{x y
0.40756 0.61687
0.19231 0.37312
0.058666 0.15181
0.018449 0.059871
0.0058375 0.023478
0.0018611 0.0087486
0.00059417 0.0028404
0.00018948 0.001144
6.0467e-05 0.00044771
1.9326e-05 0.00016855
3.6577e-06 1.8914e-05
};\addlegendentry{$(128,64)$ dynamic 5G polar, $L=1,L^\prime=2$, $2$~dB}

\addplot[gray,dashed]
table[]{x y
1e-5 1e-5
1 1
};

\end{loglogaxis}

\end{tikzpicture}
	\caption{Approximated blockwise SO vs. \ac{BLER} of polar-like codes with proposed scheme and Forney's approximation. The proposed method (solid) works with SO-SCL decoding of list size $L$, while the Forney's approximation (dashed) works with SCL decoding of list size $L^\prime$.}
	\label{fig:misdetection_rates_log}
\end{figure}

Fig.~\ref{fig:misdetection_rates_log} plots the \ac{BLER} given $\text{E}\left[1-\Gamma^*\left(y^N, \hat{u}^N\right)\right]$ and $\text{E}\left[1-\Gamma^\prime\left(y^N, \hat{u}^N\right)\right]$. The results in Fig.~\ref{fig:misdetection_rates_log} show that $\text{E}\left[1-\Gamma^*\left(y^N, \hat{u}^N\right)\right]$ accurately predicts the \ac{BLER} of the polar-like codes with dynamic frozen constraints (\ref{line:ebch32_26},\ref{line:rm64_42},\ref{line:drm128_64},\ref{line:dpolar128_64}). However, $\text{E}\left[1-\Gamma^*\left(y^N, \hat{u}^N\right)\right]$ shows a mismatch for polar-like codes with static frozen bits (\ref{line:rm128_64},\ref{line:polar128_64}). The main reason is as follows. The approximated codebook probability Eq.~\eqref{eq:APU} relies on the assumption of uniform leaf distribution under the unvisited nodes. However, the static frozen bits (after the first message bit) may disrupt this assumption. Thus, the approximation in~Eq.~\eqref{eq:core_approx} has a mismatch for polar-like codes with static frozen bits. Observe that our approximation yields accurate predictions for $(32,26)$ static RM code (\ref{line:ebch32_26}) because the number of frozen bits following the first message bit is minimal and insufficient to disrupt the assumption of a uniform leaf distribution.\footnote{Since there are only three frozen bits after the first message bit for the $(32,26)$ static RM code, the $(32,26)$ static RM code has very similar properties to the $(32,26)$ dynamic RM code.} 


To conclude, $Q_\mathcal{U}^*\left(y^N\right)$ provides an accurate approximation of the codebook probability for polar-like codes of any length and rate, if the frozen bits are random linear combinations of previous message bits, as defined in \cite[Definition 1]{cocskun2020successive}. However, $Q_\mathcal{U}^*\left(y^N\right)$ may exhibit a mismatch if the frozen constraints are static or if the constraint length of the convolutional dynamic constraints is too short.

\begin{figure}[t]
	\centering
	\begin{tikzpicture}[scale=1]
\footnotesize
\begin{loglogaxis}[
legend style={at={(1,0)},anchor= south east},
ymin=1e-5,
ymax=1,
width=3.5in,
height=3.5in,
grid=both,
xmin = 1e-5,
xmax = 1,
xlabel = {$\text{E}\left[1-\Gamma^*\left(y^N, \mathcal{L}_U\right)\right]$},
ylabel = {LER},
]

\addplot[red,mark=o]
table[]{x y
0.53635 0.53709
0.18515 0.18492
0.058802 0.058778
0.018632 0.018788
0.0059225 0.0058592
0.0018859 0.0018698
0.00060166 0.00063854
0.00019206 0.00019593
6.1295e-05 5.7403e-05
1.9605e-05 2.059e-05
};\addlegendentry{$(32,26)$ static RM, $L=2$, $3$~dB}

\addplot[blue,mark=x]
table[]{x y
0.85523 0.85842
0.18286 0.18109
0.056852 0.0579
0.018099 0.018731
0.0057299 0.0064431
0.0018352 0.002066
0.0005831 0.0005685
0.00018532 0.00019491
9.8775e-05 0.00010545
};\addlegendentry{$(64,42)$ dynamic RM, $L=2$, $3$~dB}

\addplot[brown,mark=o]
table[]{x y
0.83899 0.92614
0.1899 0.60446
0.05761 0.41207
0.018317 0.21501
0.00576 0.12441
0.0018486 0.050847
0.00057252 0.023952
0.00018531 0.010322
5.7481e-05 0.005006
1.8464e-05 0.0024636
};\addlegendentry{$(128,64)$ static RM, $L=4$, $2$~dB}

\addplot[brown,mark=x]
table[]{x y
0.96584 0.96859
0.19096 0.20541
0.056701 0.060627
0.018346 0.020407
0.0058301 0.0059846
0.0018149 0.001867
0.00057632 0.0006012
0.00018357 0.0002079
5.7767e-05 5.0021e-05
};\addlegendentry{$(128,64)$ dynamic RM, $L=4$, $2$~dB}

\addplot[gray,dashed]
table[]{x y
1e-5 1e-5
1 1
};

\end{loglogaxis}

\end{tikzpicture}
	\caption{$\text{E}\left[1-\Gamma^*\left(y^N, \mathcal{L}_U\right)\right]$ vs. \ac{LER} of polar-like codes under SO-SCL decoding with list size $L$.}
	\label{fig:list_misdetection_rates}
\end{figure}
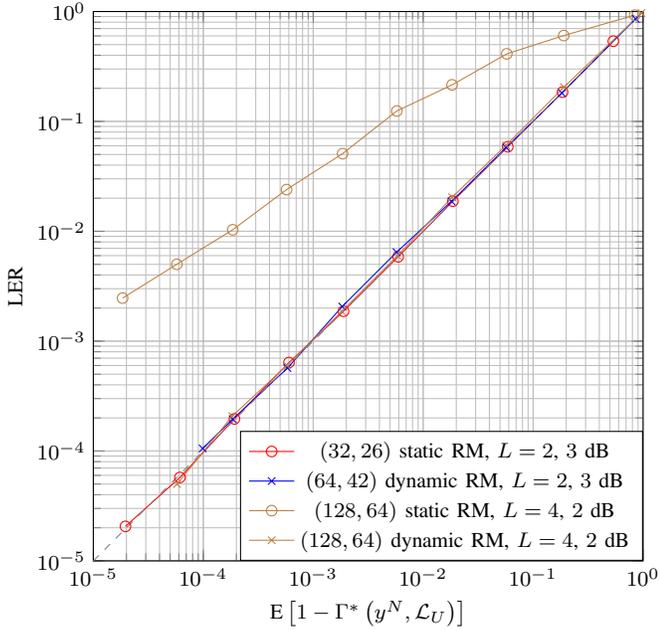

Similar to Fig.~\ref{fig:misdetection_rates_log}, Fig.~\ref{fig:list_misdetection_rates} plots the \ac{LER} given $\text{E}\left[1-\Gamma^*\left(y^N, \mathcal{L}_U\right)\right]$, where the \ac{LER} is defined as the probability of the transmitted codeword not being in the list. The results show that $\Gamma^*\left(y^N, \mathcal{L}_U\right)$ accurately predicts the \ac{LER} of the polar-like codes with dynamic frozen constraints. 

\subsection{Joint error correction and detection}
We apply a threshold test to the blockwise \ac{SO} generated by \ac{SO-SCL} decoding. The most likely candidate $\hat{u}^N$ in the list is accepted if
\begin{align*}
\Gamma^*\left(y^N, \hat{u}^N\right) > 1-\epsilon,
\end{align*}
otherwise, the decoder returns an erasure flag indicating a detected error.

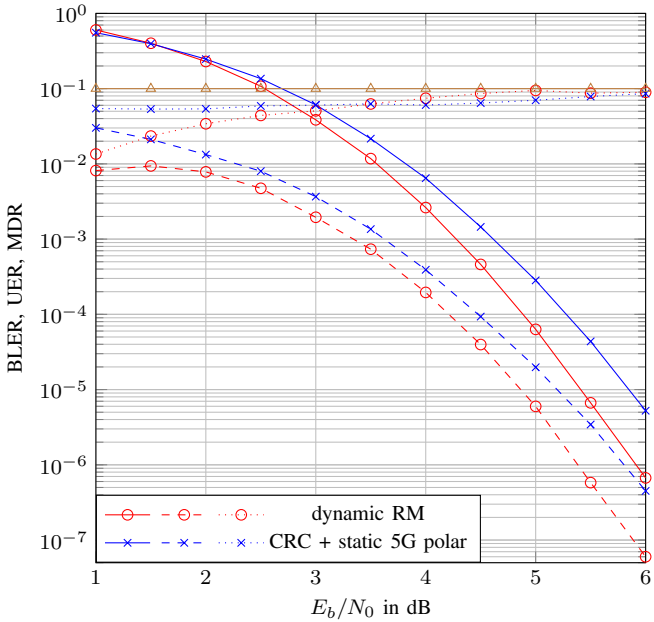
\begin{figure}[t]
	\centering
	\begin{tikzpicture}[scale=1]
\footnotesize
\begin{semilogyaxis}[
legend style={at={(0,0)},anchor= south west},
legend columns=3,
ymin=5e-8,
ymax=1,
width=3.5in,
height=3.5in,
grid=both,
xmin = 1,
xmax = 6,
xlabel = $E_b/N_0$ in dB,
ylabel = {BLER, UER, MDR},
]

\addplot[red,mark=o]
table[]{x y
1 0.60241
1.5 0.40115
2 0.22936
2.5 0.10753
3 0.038509
3.5 0.011744
4 0.0026254
4.5 0.00046061
5 6.3136e-05
5.5 6.66e-06
6 6.70e-07
};\addlegendentry{ }

\addplot[red,mark=o,dashed,mark options=solid]
table[]{x y
1 0.0081446
1.5 0.0093946
2 0.0078269
2.5 0.0047351
3 0.0019508
3.5 0.00073438
4 0.00019558
4.5 3.968e-05
5 5.9788e-06
5.5 5.8e-07
6 6.0e-08
};\addlegendentry{ }

\addplot[red,mark=o,dotted,mark options=solid]
table[]{x y
1 0.01352
1.5 0.023419
2 0.034125
2.5 0.044037
3 0.050659
3.5 0.062531
4 0.074493
4.5 0.086147
5 0.094697
5.5 0.087087
6 0.089552
};\addlegendentry{dynamic RM}

\addplot[blue,mark=x]
table[]{x y
1 0.5539
1.5 0.39582
2 0.24677
2.5 0.1358
3 0.060869
3.5 0.021622
4 0.0064603
4.5 0.001452
5 0.00028272
5.5 4.3683e-05
6 5.24e-06
};\addlegendentry{ }

\addplot[blue,mark=x,dashed,mark options=solid]
table[]{x y
1 0.0301
1.5 0.021158
2 0.013284
2.5 0.0079921
3 0.0036721
3.5 0.0013575
4 0.00039116
4.5 9.3366e-05
5 1.981e-05
5.5 3.4331e-06
6 4.5e-07
};\addlegendentry{ }

\addplot[blue,mark=x,dotted,mark options=solid]
table[]{x y
1 0.054342
1.5 0.053453
2 0.053833
2.5 0.058851
3 0.060328
3.5 0.062783
4 0.060547
4.5 0.0643
5 0.070067
5.5 0.078592
6 0.085878
};\addlegendentry{CRC + static 5G polar}





\addplot[brown,mark=triangle]
table[]{x y
1 0.1
1.5 0.1
2 0.1
2.5 0.1
3 0.1
3.5 0.1
4 0.1
4.5 0.1
5 0.1
5.5 0.1
6 0.1
};

\end{semilogyaxis}

\end{tikzpicture}
	\caption{BLER(solid), UER(dashed), MDR(dotted) vs. $E_b/N_0$ over the biAWGN channel for the $\left(64,42\right)$ dynamic RM code compared to a $(64,42+6)$ static 5G polar code with an outer CRC-$6$ \texttt{0x30}. SCL with $L=4$, threshold $\epsilon=0.1$}
	\label{fig:uBLER1}
\end{figure}

\begin{figure}[t]
	\centering
	\begin{tikzpicture}[scale=1]
\footnotesize
\begin{semilogyaxis}[
legend style={at={(0,0)},anchor= south west},
legend columns=3,
ymin=1e-8,
ymax=1,
width=3.5in,
height=3.5in,
grid=both,
xmin = 1,
xmax = 6,
xlabel = $E_b/N_0$ in dB,
ylabel = {BLER, UER, MDR},
]

\addplot[red,mark=o]
table[]{x y
1 0.83181
1.5 0.65583
2 0.43712
2.5 0.23675
3 0.10013
3.5 0.032457
4 0.0079574
4.5 0.0015206
5 0.00022239
5.5 2.642e-05
6 2.35e-06
};\addlegendentry{ }

\addplot[red,mark=o,dashed,mark options=solid]
table[]{x y
1 0.00024961
1.5 0.00037476
2 0.0004196
2.5 0.00038787
3 0.00019598
3.5 7.8429e-05
4 2.3646e-05
4.5 5.954e-06
5 9.9e-07
5.5 1.1e-07
6 1.0e-08
};\addlegendentry{ }

\addplot[red,mark=o,dotted,mark options=solid]
table[]{x y
1 0.00030008
1.5 0.00057143
2 0.00095991
2.5 0.0016383
3 0.0019572
3.5 0.0024164
4 0.0029716
4.5 0.0039155
5 0.0044516
5.5 0.0041635
6 0.0042553
};\addlegendentry{dynamic RM}

\addplot[blue,mark=x]
table[]{x y
1 0.7599
1.5 0.62689
2 0.47078
2.5 0.3097
3 0.17369
3.5 0.081405
4 0.030948
4.5 0.009098
5 0.0020594
5.5 0.00034859
6 4.289e-05
};\addlegendentry{ }

\addplot[blue,mark=x,dashed,mark options=solid]
table[]{x y
1 0.0026286
1.5 0.0020657
2 0.0020387
2.5 0.0011611
3 0.00067075
3.5 0.00031006
4 0.00014226
4.5 4.8698e-05
5 1.3665e-05
5.5 2.8633e-06
6 4.9e-07
};\addlegendentry{ }

\addplot[blue,mark=x,dotted,mark options=solid]
table[]{x y
1 0.0034592
1.5 0.0032951
2 0.0043304
2.5 0.0037492
3 0.0038618
3.5 0.0038089
4 0.0045967
4.5 0.0053526
5 0.0066355
5.5 0.008214
6 0.011425
};\addlegendentry{CRC+ static 5G polar}

\addplot[brown,mark=triangle]
table[]{x y
1 0.005
1.5 0.005
2 0.005
2.5 0.005
3 0.005
3.5 0.005
4 0.005
4.5 0.005
5 0.005
5.5 0.005
6 0.005
};\label{line:designMDR}

\end{semilogyaxis}
   
\end{tikzpicture}
	\caption{BLER(solid), UER(dashed), MDR(dotted) vs. $E_b/N_0$ over the biAWGN channel for the $\left(64,42\right)$ dynamic RM code compared to a $(64,42+11)$ static 5G polar code with an outer CRC-$11$ \texttt{0x710}. SCL with $L=8$, threshold $\epsilon=0.005$}
	\label{fig:uBLER2}
\end{figure}
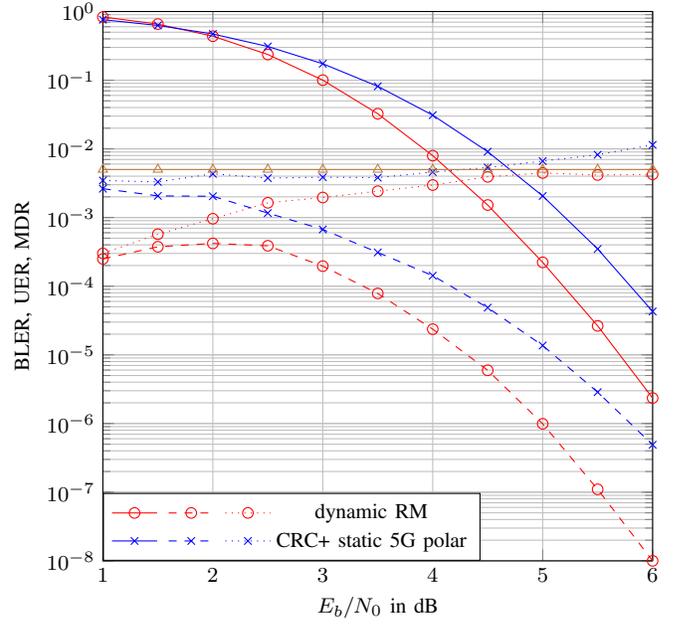

Fig.~\ref{fig:uBLER1} and Fig.~\ref{fig:uBLER2} show the \ac{BLER}, \ac{UER} and \ac{MDR} of the proposed decision rule based on \ac{SO-SCL}. For reference, we demonstrate the performance of CRC-concatenated polar codes using \ac{SCL}~\cite{tal2015list,niu2012crc} with the same list size. If none of the candidates in the list pass the CRC, an erasure flag is returned, indicating that an error has been detected. Our method is compared with $6$ bits CRC using $L=4$ and $11$ bits CRC using $L=8$. The threshold $\epsilon$ is chosen to achieve a similar \ac{MDR} to that of CRC-concatenated polar codes. The simulation results show that the proposed method outperforms CRC-concatenated polar codes in both \ac{BLER} and \ac{UER}. More importantly, the \ac{MDR} of \ac{SO-SCL} is restricted to not be higher than $\epsilon$ (the horizontal line \ref{line:designMDR} in Fig.~\ref{fig:uBLER1} and Fig.~\ref{fig:uBLER2}).\footnote{Note that the \ac{MDR} of CRC-concatenated polar codes using \ac{SCL} is influenced by both the CRC size and the list size. The generator polynomial is presented with Koopman's notation~\cite{koopman_crc}.}

\subsection{Turbo decoding of product codes and GLDPC codes}
As explained in Sec.\ref{sec:SISO}, \ac{APP} \acp{LLR} can be approximated from a candidate list. The list-sum approximation in Eq.~\eqref{eq:list-sum}, based on \ac{SCL} decoding, was introduced in~\cite{liu2017parallel} for parallel concatenated polar codes. In~\cite{bioglio2019construction,condo2020practical,cocskun2024precoded}, the list-max approximation in Eq.~\eqref{eq:list-max} is used to decode polar product codes with \ac{SCL} decoding.

In this section, we show the comparison between the turbo decoder based on \ac{SCL} with Pyndiah's list-max approximation Eq.~\eqref{eq:list-max}, and the turbo decoder based on \ac{SO-SCL} Eq.~\eqref{eq:approx_app}. Both turbo decoders have a maximum iteration number of $I_\text{max}=20$. For both turbo decoders, all component codes are initially decoded by an \ac{SCL} decoder with list size of $L=4$. Then, \enquote{SCL, list-max} and SO-SCL extract the bitwise \ac{SO} by using Eq.~\eqref{eq:list-max} and Eq.~\eqref{eq:approx_app}, respectively. 

\begin{figure}[t]
	\centering
	\begin{tikzpicture}[scale=1]
\footnotesize
\begin{semilogyaxis}[
legend style={at={(0.02,0.02)},anchor= south west},
legend columns=1,
ymin=1e-7,
ymax=1,
width=3.5in,
height=3.5in,
grid=both,
xmin = 1,
xmax = 2.25,
xlabel = $E_b/N_0$ in dB,
ylabel = {BLER, BER},
]

\addplot[blue,mark=x]
table[]{x y
1 0.9434
1.25 0.75758
1.5 0.27027
1.75 0.021295
2 0.0003672
2.25 1.0208e-06
};\addlegendentry{SCL, list-max}

\addplot[red,mark=o]
table[]{x y
1 0.42373
1.25 0.04995
1.5 0.001197
1.75 9e-06
2 0
};\addlegendentry{SO-SCL}

\addplot[blue,mark=x,dashed,mark options=solid]
table[]{x y
1 0.21011
1.25 0.14621
1.5 0.060905
1.75 0.0044934
2 7.6061e-05
2.25 2.124e-07
2.5 0
};

\addplot[red,mark=o,dashed,mark options=solid]
table[]{x y
1 0.05361
1.25 0.0058086
1.5 0.00014459
1.75 9.9146e-07
2 0
};

\end{semilogyaxis}

\end{tikzpicture}
	\caption{BLER(solid), BER(dashed) vs. $E_b/N_0$ over the biAWGN channel for the $\left(4096,1764\right)$ product code based on $\left(64,42\right)$ dynamic RM codes.}
	\label{fig:prod_polar_64_42}
\end{figure}
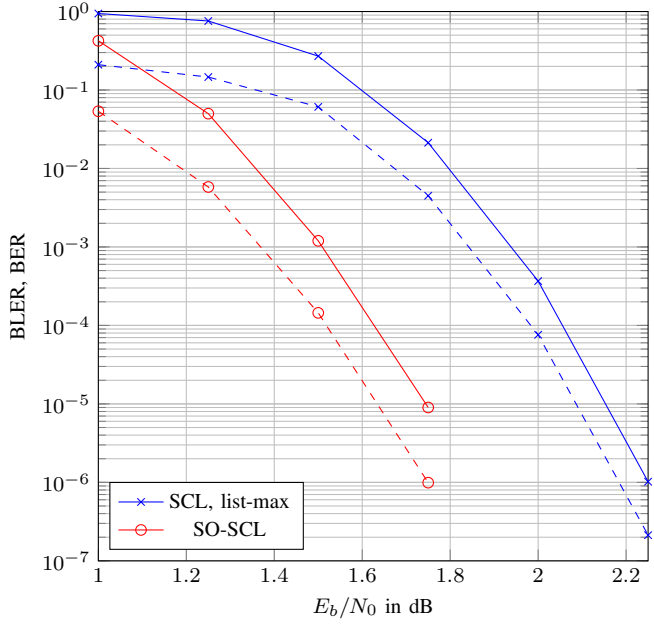

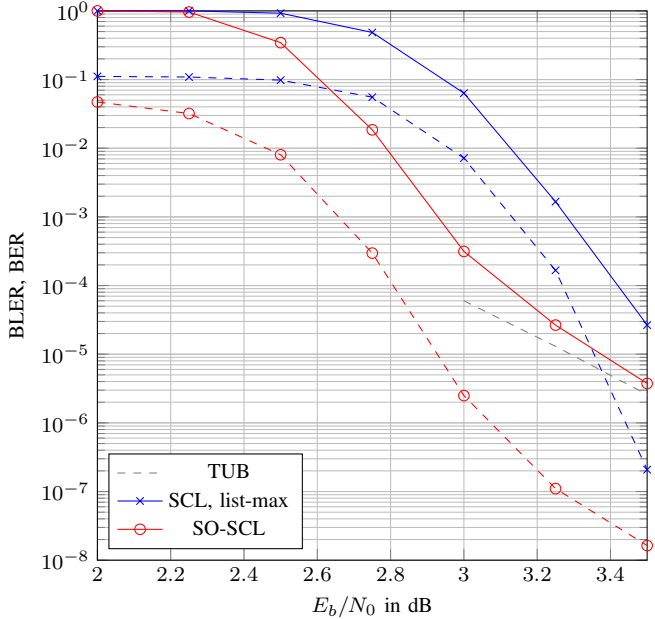
\begin{figure}[t]
	\centering
	\begin{tikzpicture}[scale=1]
\footnotesize
\begin{semilogyaxis}[
legend style={at={(0.02,0.02)},anchor= south west},
legend columns=1,
ymin=1e-8,
ymax=1,
width=3.5in,
height=3.5in,
grid=both,
xmin = 2,
xmax = 3.5,
xlabel = $E_b/N_0$ in dB,
ylabel = {BLER, BER},
]

\addplot[gray,dashed]
table[]{x y
3 6.0028e-05
3.25 1.3021e-05
3.5 2.5841e-06
3.75 4.6671e-07
4 7.6286e-08
};\addlegendentry{TUB}

\addplot[blue,mark=x]
table[]{x y
2 1
2.25 1
2.5 0.92593
2.75 0.48544
3 0.063371
3.25 0.0016652
3.5 2.6579e-05
};\addlegendentry{SCL, list-max}

\addplot[red,mark=o]
table[]{x y
2 1
2.25 0.96154
2.5 0.34483
2.75 0.018525
3 0.00031403
3.25 2.6543e-05
3.5 3.7465e-06
};\addlegendentry{SO-SCL}

\addplot[blue,mark=x,dashed,mark options=solid]
table[]{x y
2 0.11111
2.25 0.10888
2.5 0.098098
2.75 0.055399
3 0.0071821
3.25 0.00016653
3.5 2.0801e-07
};

\addplot[red,mark=o,dashed,mark options=solid]
table[]{x y
2 0.047144
2.25 0.032009
2.5 0.0080285
2.75 0.00029659
3 2.4958e-06
3.25 1.1039e-07
3.5 1.6344e-08
};

\end{semilogyaxis}

\end{tikzpicture}
	\caption{BLER(solid), BER(dashed) vs. $E_b/N_0$ over the biAWGN channel for the $\left(4096,3249\right)$ product code based on $\left(64,57\right)$ static RM codes.}
	\label{fig:prod_polar_64_57}
\end{figure}

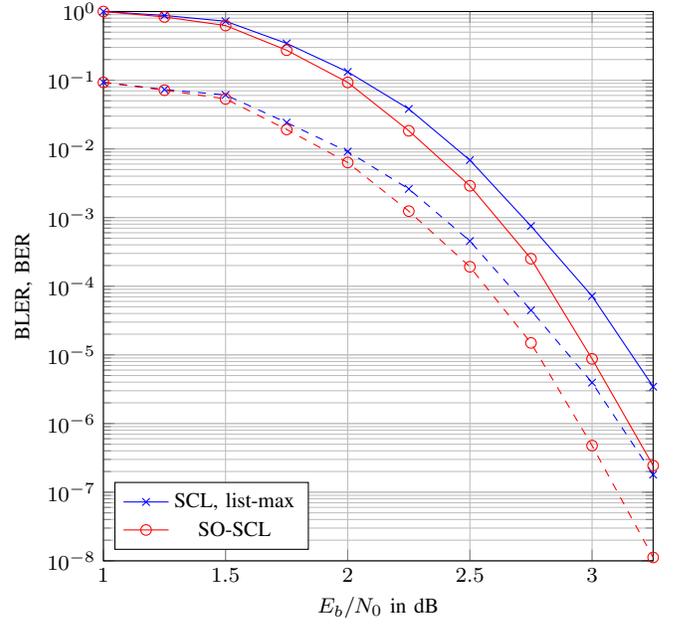
\begin{figure}[t]
	\centering
	\begin{tikzpicture}[scale=1]
\footnotesize
\begin{semilogyaxis}[
legend style={at={(0.02,0.02)},anchor= south west},
legend columns=1,
ymin=1e-8,
ymax=1,
width=3.5in,
height=3.5in,
grid=both,
xmin = 1,
xmax = 3.25,
xlabel = $E_b/N_0$ in dB,
ylabel = {BLER, BER},
]

\addplot[blue,mark=x]
table[]{x y
1 1
1.25 0.8750
1.5 0.7221
1.75 0.34321
2 0.13222
2.25 0.038122
2.5 0.0068542
2.75 0.00075111
3 7.1786e-05
3.25 3.4352e-06
};\addlegendentry{SCL, list-max}

\addplot[red,mark=o]
table[]{x y
1 1
1.25 0.83333
1.5 0.625
1.75 0.27322
2 0.09285
2.25 0.018335
2.5 0.0028964
2.75 0.00025167
3 8.7361e-06
3.25 2.4352e-07
};\addlegendentry{SO-SCL}

\addplot[blue,mark=x,dashed,mark options=solid]
table[]{x y
1 0.09322
1.25 0.073011
1.5 0.06099
1.75 0.024122
2 0.0091255
2.25 0.0026088
2.5 0.00045402
2.75 4.4598e-05
3 3.9452e-06
3.25 1.8018e-07
};

\addplot[red,mark=o,dashed,mark options=solid]
table[]{x y
1 0.092734
1.25 0.071517
1.5 0.053369
1.75 0.019152
2 0.0063284
2.25 0.0012373
2.5 0.00019183
2.75 1.4899e-05
3 4.769e-07
3.25 1.121e-08
};

\end{semilogyaxis}

\end{tikzpicture}
	\caption{BLER(solid), BER(dashed) vs. $E_b/N_0$ over the biAWGN channel for the $\left(1024,640\right)$ GLDPC code based on $\left(32,26\right)$ static RM check nodes.}
	\label{fig:gldpc_polar_32_26}
\end{figure}

In Fig.~\ref{fig:prod_polar_32_26},\ref{fig:prod_polar_64_42},\ref{fig:prod_polar_64_57}, we demonstrate the \ac{BLER} and \ac{BER} of $(32^2,26^2)$, $(64^2,42^2)$ and $(64^2,57^2)$ product codes based on static/dynamic \ac{RM} component codes. The \acp{TUB} of $(32^2,26^2)$ and $(64^2,57^2)$ codes are provided. In Fig.~\ref{fig:gldpc_polar_32_26}, we show the performance of a $(1024,640)$ \ac{GLDPC} code based on $(32,26)$ static \ac{RM} check nodes introduced in~\cite{Lentmaier10}. For \enquote{SCL, list-max}, $\alpha$ and $\beta$ parameters are iteration-dependent taken from~\cite{pyndiah_1998}. For \ac{SO-SCL}, the extrinsic \acp{LLR} are always scaled by $0.5$ for all product codes and $0.6$ for the \ac{GLDPC} code. 

Simulation results show that the turbo decoder with proposed \ac{SO-SCL} significantly outperforms that with list-max approximation. For the high-rate $(32^2,26^2)$ and $(64^2,57^2)$ product codes, the performance is close to their \acp{TUB} with \ac{SO-SCL} of list size $4$. Note that the component polar-like codes are systematically encoded~\cite{arikan2011systematic} to reduce the \ac{BER}. 

\begin{figure}[t]
	\centering
	\begin{tikzpicture}[scale=1]
\footnotesize
\begin{semilogyaxis}[
legend style={at={(0.02,0.02)},anchor= south west},
legend columns=1,
ymin=1e-8,
ymax=1,
width=3.5in,
height=3.5in,
grid=both,
xmin = 1,
xmax = 3.5,
xlabel = $E_b/N_0$ in dB,
ylabel = {BLER, BER},
]

\addplot[gray,dashed]
table[]{x y
2.5 0.00067935
2.75 0.00021718
3 6.4992e-05
3.25 1.8136e-05
3.5 4.6997e-06
};\addlegendentry{TUB}

\addplot[blue,mark=x]
table[]{x y
1 0.96154
1.25 0.9434
1.5 0.61728
1.75 0.34965
2 0.15152
2.25 0.04095
2.5 0.0065334
2.75 0.0012417
3 0.00023035
3.25 4.6753e-05
3.5 1e-05
};\addlegendentry{SCL, list-max}

\addplot[red,mark=o]
table[]{x y
1 0.96154
1.25 0.81967
1.5 0.48077
1.75 0.22222
2 0.048544
2.25 0.0075873
2.5 0.0018201
2.75 0.00038972
3 0.00009923
3.25 2.3825e-05
3.5 5e-06
};\addlegendentry{SO-SCL}

\addplot[blue,mark=x,dashed,mark options=solid]
table[]{x y
1 0.15445
1.25 0.14534
1.5 0.089844
1.75 0.047216
2 0.018649
2.25 0.0049364
2.5 0.00064007
2.75 6.4268e-05
3 7.5403e-06
3.25 9.045e-07
3.5 1.25e-07
};

\addplot[red,mark=o,dashed,mark options=solid]
table[]{x y
1 0.077562
1.25 0.061675
1.5 0.029154
1.75 0.0098187
2 0.0023364
2.25 0.00028206
2.5 3.5393e-05
2.75 8.0644e-06
3 1.755e-06
3.25 3.7391e-07
3.5 7.8125e-08
};

\end{semilogyaxis}

\end{tikzpicture}
	\caption{BLER(solid), BER(dashed) vs. $E_b/N_0$ over the biAWGN channel for the $\left(1024,676\right)$ product code based on $\left(32,26\right)$ static RM codes.}
	\label{fig:prod_polar_32_26}
\end{figure}
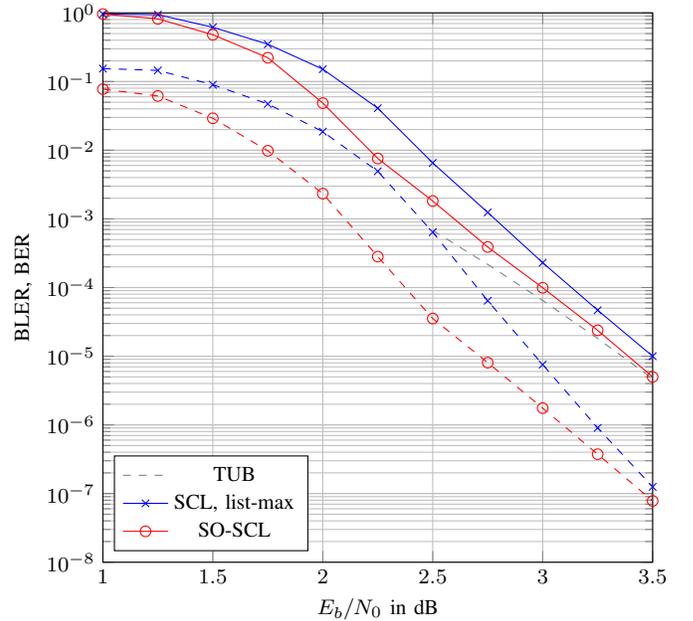

\subsection{Quality of the bitwise SO}
In this section, we evaluate the quality of the bitwise \ac{SO} generated by \ac{SO-SCL} Eq.~\eqref{eq:approx_app}. To the best of the authors' knowledge, no information-theoretically motivated evaluation of \ac{SISO} decoders has been reported to date. In this work, we demonstrate the quality of bitwise \ac{SO} by analyzing: 1) \ac{BER}, 2) \ac{GMI}, and 3) the ensemble iterative decoding threshold. We compare \ac{SO-SCL} with the list-max approximation Eq.~\eqref{eq:list-max}, the list-sum approximation Eq.~\eqref{eq:list-sum}, and the \ac{BCJR} decoder~\cite{bahl1974optimal}.

\subsubsection*{Bit Error Rate}
A bitwise \ac{MAP} decoder, defined as
\begin{align*}
\hat{c}_{\text{MAP},i} \triangleq \argmax_{a\in\{0,1\}} P_{C_i|Y^N}\left(a\left|y^N\right.\right)
\end{align*}
which can be implemented by performing a hard decision on the \ac{APP} \acp{LLR} produced by the \ac{BCJR} algorithm. 
\begin{align*}
\hat{c}_{\text{MAP},i} =\left\{
    \begin{aligned}
      &0, \text{ if } \ell_{\text{APP},i}\geq 0\\
      &1,\text{ if } \ell_{\text{APP},i}< 0.
    \end{aligned}\right.
\end{align*}
The \ac{MAP} decoder provides an optimal \ac{BER} by its definition. Higher quality bitwise \ac{SO} should result in a lower \ac{BER}.

Tab.~\ref{tab:BER} shows the \ac{BER} of the $(32,26)$ static \ac{RM} code by performing a hard decision on the bitwise \ac{SO} generated by \ac{SO-SCL} Eq.~\eqref{eq:approx_app}, list-max approximation Eq.~\eqref{eq:list-max} and list-sum approximation Eq.~\eqref{eq:list-sum}. For reference, the \ac{BER} of \ac{SCL} is also provided. Simulation results show that the \ac{SO-SCL} performs closer to the optimal \ac{MAP} decoder than other approximations.

\begin{table*}[t]
    \centering
    \footnotesize
    \setlength{\tabcolsep}{10pt}
    \renewcommand\arraystretch{1.5}
    \caption{BER of $(32,26)$ static RM codes with different SISO decoders.}
\begin{tabular}{ |c|c|c|c|c|c|c|c| }
\hline
 \multicolumn{2}{| c |}{$E_b/N_0$} & $0$ dB & $1$ dB & $2$ dB & $3$ dB & $4$ dB & $5$ dB\\
\hline
\hline
\multirow{5}{*}{\shortstack{BER}} & SO-SCL ($L=4$) & $0.093712$  & $0.061758$ & $0.032229$ & $0.011629$ & $0.0027179$&$0.00037739$\\ 
\cline{2-8}
 &list-max ($L=4$) & $0.102810$ & $0.066558$& $0.034110$ & $0.011980$ & $0.0027626$&$0.00038120$\\ 
\cline{2-8}
 &list-sum ($L=4$) & $0.097503$ & $0.063655$& $0.032938$ &$0.011685$ &$0.0027188$ &$0.00037792$\\
\cline{2-8}
&SCL ($L=4$) & $0.102810$ & $0.066558$& $0.034110$ &$0.011980$ &$0.0027626$ &$0.00038120$\\
\cline{2-8}
 &MAP & $0.093537$ & $0.061489$& $0.032149$  & $0.011582$ &$0.0027080$&$0.00037664$\\
\hline
\end{tabular}
    \label{tab:BER}
\end{table*}

\subsubsection*{Generalized Mutual Information}
In systems with iterative processing, a SISO decoder of an $(N,K)$ code works as a generalized \ac{CN} with $N$ edges.
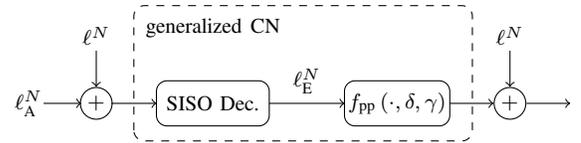
\begin{figure}[t]
	\centering
	\begin{tikzpicture}[scale=1]
    \footnotesize

\node at (-1.7,0.3) {$\ell_{\text{A}}^N$};

\draw [->] (-0.8,1)--(-0.8,0.52);
\node at (-0.8,1.2) {$\ell^N$};

\draw [->] (-1.5,0.3)--(-1,0.3);
\draw [] (-0.8,0.3) circle (6pt);
\node at (-0.8,0.3) {{$+$}};
    
\draw[rounded corners] (0,0) rectangle (1.5,0.6);
\node at (0.75,0.3) {SISO Dec.};

\draw [->] (-0.6,0.3)--(0,0.3);

\node at (2,0.6) {$\ell_{\text{E}}^N$};
\draw [->] (1.5,0.3)--(2.5,0.3);
\draw[rounded corners] (2.5,0) rectangle (3.9,0.6);
\node at (3.2,0.3) {$f_\text{pp}\left(\cdot,\delta,\gamma\right)$};

\draw [->] (3.9,0.3)--(4.48,0.3);
\draw [] (4.7,0.3) circle (6pt);
\node at (4.7,0.3) {{$+$}};

\draw [->] (4.7,1)--(4.7,0.52);
\node at (4.7,1.2) {$\ell^N$};

\draw [->] (4.92,0.3)--(5.5,0.3);

\draw[dashed, rounded corners] (-0.3,-0.2) rectangle (4.2,1.6);

\node at (0.75,1.3) {generalized CN};

\end{tikzpicture}
	\caption{Generalized \ac{CN} based on an \ac{SISO} decoder.}
	\label{fig:G-CN}
\end{figure}
The generalized \ac{CN} (Fig.~\ref{fig:G-CN}) takes the channel observation and the a-priori information as input and the outputs extrinsic information. Due to the imperfection of the \ac{SISO} decoder and the specific structure of the system, a post-processing is required, e.g., Pyndiah introduced iteration-dependent parameters $\alpha,\beta$~\cite{pyndiah_1998} for list-max approximation. In~\cite{strasshofer2023soft}, an information-theoretically motivated method is proposed to optimize the post-processing,
\begin{align*}
f_\text{pp}\left(\ell_{\text{E}},\delta,\gamma\right) =\left\{
    \begin{aligned}
&\gamma\text{sign}\left(\ell_{\text{E}}\right),&\text{ if } \ell_{\text{E}} = \pm\infty \\
      &\delta\ell_{\text{E}},&\text{ otherwise } 
    \end{aligned}\right.
\end{align*}
which includes scaling and saturation for $\pm\infty$, such that the $1$-GMI~\cite{kaplan1993information} is maximized. Note that \ac{SO-SCL} and \ac{BCJR} do not necessitate a saturation value.

The generalized \ac{CN} updates the bitwise \acp{LLR} from $\ell_\text{ch}^N+\ell_{\text{A}}^N$ to $\ell_\text{ch}^N+f_\text{pp}\left(\ell_{\text{E}}^N,\delta,\gamma\right)$ for the next soft-input device. We use $1$-GMI to describe the quality of the bitwise \ac{SO}, i.e.,
\begin{align*}
&I_1\left(C;L_{\text{ch}}+f_\text{pp}\left(L_{\text{E}},\delta,\gamma\right)\right)\\
=&~1-\text{E}\left[\log_2\left(1+e^{-\left(1-2\cdot C\right)\cdot\left(L_{\text{ch}}+f_\text{pp}\left(L_{\text{E}},\delta,\gamma\right)\right)}\right)\right].
\end{align*}
The simulation is designed as follows. A large number ($10^6$) of codewords are transmitted over a \ac{biAWGN} channel. Assume that we do not have any a-priori information yet, i.e., $\ell_\text{A}^N=0^N$. The SISO decoder takes only channel \acp{LLR} $\ell_\text{ch}^N$ as input and outputs the extrinsic \acp{LLR} $\ell_\text{E}^N$. We evaluate the $1$-GMI with optimal post-processing after the decoding,
\begin{align}\label{eq:optimal_pp}
\max_{\delta,\gamma}I_1\left(C;L_{\text{ch}}+f_\text{pp}\left(L_{\text{E}},\delta,\gamma\right)\right)
\end{align}
which describes the correlation between transmitted codeword $c^N$ and the \acp{LLR} $\ell_\text{ch}^N+f_\text{pp}\left(\ell_{\text{E}}^N,\delta,\gamma\right)$ delivered to the next soft-input device. The simulation results for the $(32,26)$ static RM code are demonstrated in Tab.~\ref{tab:GMI}. We observe that the \ac{SO-SCL} provides higher $1$-GMI than other approximations.

\begin{table*}[t]
    \centering
    \footnotesize
    \setlength{\tabcolsep}{10pt}
    \renewcommand\arraystretch{1.5}
    \caption{$1$-GMI of $(32,26)$ static RM codes with different SISO decoders.}
\begin{tabular}{ |c|c|c|c|c|c|c|c| }
\hline
 \multicolumn{2}{| c |}{$E_b/N_0$} & $0$ dB & $1$ dB & $2$ dB & $3$ dB & $4$ dB & $5$ dB\\
\hline
\hline
\multirow{4}{*}{\shortstack{$1$-GMI}} & SO-SCL ($L=4$) & $0.66733$& $0.76970$& $0.87071$ & $0.95215$ & $0.98799$ &$0.99838$\\ 
\cline{2-8}
&list-max ($L=4$) & $0.66553$ & $0.76758$&$0.86867$ &$0.95092$ & $0.98767$&$0.99834$\\ 
\cline{2-8}
&list-sum ($L=4$) & $0.66452$& $0.76677$&  $0.86838$& $0.95109$ & $0.98772$&$0.99836$\\ 
\cline{2-8}
&BCJR & $0.67231$ & $0.77658$& $0.87690$ & $0.95403$ & $0.98829$&$0.99842$\\
\hline
\end{tabular}
    \label{tab:GMI}
\end{table*}

\subsubsection*{Ensemble Iterative Decoding Threshold}
In this section, we analyze the iterative decoding threshold of the turbo-like ensembles~\cite{moloudi2017spatially}. We consider a regular \ac{GLDPC} code ensemble with \ac{VN} degree $2$ and all generalized \acp{CN} are $\left(N,K\right)$ static RM check nodes~\cite{liva2008quasi,Lentmaier10}. Assume that we have $2M$ generalized \acp{CN}, each of which is connected to $N$ \acp{VN}. Each \ac{VN} is part of the constraints of $2$ generalized \acp{CN}. The code length is $MN$, and there are a total of $2M(N-K)$ constraints. The ensemble code rate is given by 
\begin{align*}
R=\frac{MN-2M(N-K)}{MN}=\frac{2K-N}{N}
\end{align*}
if all of the constraints derived from the generalized \acp{CN} are linearly independent.

We find the iterative decoding threshold via \ac{MCDE}~\cite{jian2012approaching,sheikh2021refined} as follows. We construct a sufficiently long code with $M=10^5$. Assume that the all-zero codeword of length $MN$ is transmitted over the \ac{biAWGN} channel. We track the empirical distribution of the a-priori \acp{LLR} and the distribution of the post-processed extrinsic \acp{LLR} throughout the iterations. The ensemble iterative decoding threshold $(E_b/N_0)^*$ is defined as the lowest $E_b/N_0$ for which \enquote{the post-processed extrinsic \acp{LLR} are all larger than zero} as the number of iterations grows large. The post-processing for each iteration is individually optimized by maximizing the $1$-GMI Eq.~\eqref{eq:optimal_pp} as in~\cite{strasshofer2023soft}. The a-priori \acp{LLR} and channel \acp{LLR} are permuted before each iteration to mitigate any dependencies introduced in previous iterations~\cite{mackay1999good}. 

In Tab.~\ref{tab:threshold1} and Tab.~\ref{tab:threshold2}, we present the iterative decoding threshold $(E_b/N_0)^*$ of the previously mentioned GLDPC code ensembles with different \ac{SISO} decoders. The post-processing for each decoder and each iteration is individually optimized via Eq.~\eqref{eq:optimal_pp}. We observe that iterative decoding with \ac{SO-SCL} has a lower $(E_b/N_0)^*$ than with other approximations.

\begin{table*}[t]
    \centering
    \footnotesize
    \setlength{\tabcolsep}{10pt}
    \renewcommand\arraystretch{1.5}
    \caption{Iterative decoding thresholds for the GLDPC code ensemble with $(32, 26)$ static RM component codes.}
\begin{tabular}{ |c|c|c|c|c|c|c|c|c|c|c| }
\hline
Algorithm & BCJR& \multicolumn{3}{| c |}{SO-SCL $(L=2/4/8)$} & \multicolumn{3}{| c |}{list-max $(L=2/4/8)$} &\multicolumn{3}{| c |}{list-sum $(L=2/4/8)$}\\
\hline
$(E_b/N_0)^*$ & $1.48$ & $1.69$ &$1.58$ & $1.52$ & $2.26$ &$1.88$& $1.65$ & $2.26$ &$1.86$ & $1.60$ \\
\hline
\end{tabular}
    \label{tab:threshold1}
\end{table*}
\begin{table*}[t]
    \centering
    \footnotesize
    \setlength{\tabcolsep}{10pt}
    \renewcommand\arraystretch{1.5}
    \caption{Iterative decoding thresholds for the GLDPC code ensemble with $(64, 57)$ static RM component codes.}
\begin{tabular}{ |c|c|c|c|c|c|c|c|c|c|c| }
\hline
Algorithm & BCJR & \multicolumn{3}{| c |}{SO-SCL $(L=2/4/8)$} & \multicolumn{3}{| c |}{list-max $(L=2/4/8)$} &\multicolumn{3}{| c |}{list-sum $(L=2/4/8)$}\\
\hline
$(E_b/N_0)^*$ & $2.31$ & $2.68$&$2.54$ & $2.44$ & $3.10$&$2.76$& $2.54$ & $3.10$&$2.75$ & $2.50$ \\
\hline
\end{tabular}
    \label{tab:threshold2}
\end{table*}

\section{Conclusions}\label{sec:conclusions}
In this work, we proposed a method to approximate the codebook probability of polar-like codes based on SCL decoding. Building upon this codebook probability, we introduced SO-SCL to generate both blockwise \ac{SO} and bitwise \ac{SO}.

Simulation results indicate that the blockwise \ac{SO} accurately matches the probability of the output decision being the transmitted codeword. Dynamic \ac{RM} codes using generalized decoding that utilizes blockwise \ac{SO} significantly outperform CRC-concatenated polar codes using \ac{SCL} decoding in terms of both \ac{BLER} and \ac{UER}. More importantly, the \ac{MDR} can be constrained to not exceed any predefined value.

To enhance the accuracy of bitwise \ac{SO}, SO-SCL introduces an additional term based on the codebook probability to dynamically adjust the weight between list observation and channel observation. Both the simulation results of iterative decoding for product and GLDPC codes, as well as the information-theoretical analysis, highlight the superiority of SO-SCL over list-max and list-sum approximations.

\bibliographystyle{IEEEtran}
\bibliography{reference}

\end{document}